\documentclass[10pt,sigconf]{acmart}

\renewcommand\footnotetextcopyrightpermission[1]{} 
\setcopyright{acmlicensed}

\acmDOI{}

\acmISBN{}

\acmConference[Manuscript under review]{}  
\acmYear{2026}
\copyrightyear{2026}

\acmPrice{}

\usepackage{algorithm}
\usepackage{diagbox} 
\usepackage{algpseudocode}
\usepackage{booktabs} 
\usepackage{caption}
\usepackage{threeparttable}
\usepackage{subcaption}
\usepackage{enumerate}
\usepackage{multirow}
\usepackage{wasysym}
\usepackage{listings}
\usepackage{xcolor}
\usepackage{graphicx}
\usepackage{csquotes}
\usepackage{balance}
\usepackage{blindtext} 
\usepackage{cleveref}
\usepackage{xcolor, pifont}
\usepackage[normalem]{ulem}
\usepackage{totpages}
\usepackage{ulem}
\let\oldsout\sout
\renewcommand{\sout}[1]{\textcolor{red}{\oldsout{#1}}}

\crefformat{footnote}{#2\footnotemark[#1]#3}

\definecolor{dark green}{RGB}{94,145,40}

\newcommand{\highlight}[1]{}

\providecommand{\eg}{\emph{e.g.,} }

\newcommand{\fullname}{{\em Path Consistency Scoring}}
\newcommand{\name}{{\em PCS}}

\author{Santiago Klein}
\affiliation{
    \institution{Northwestern University}
    \country{Evanston, IL, USA}
}
\email{santiagoklein@u.northwestern.edu}

\author{Caleb J. Wang}
\affiliation{
    \institution{Northwestern University}
    \country{Evanston, IL, USA}
}
\email{caleb.wang@northwestern.edu}

\author{Fabián E. Bustamante }
\affiliation{
    \institution{Northwestern University}
    \country{Evanston, IL, USA}
}
\email{fabianb@cs.northwestern.edu}

\begin{document}

\setcounter{page}{1}       

\title[Overconfident Coordinates]{Overconfident Coordinates: Quantifying Confidence in Traceroute Geolocation}

\begin{abstract}
Studies of Internet paths often attach router locations to traceroute hops
using commercial geolocation databases, rDNS labels, Geofeeds, and IXP
metadata. These sources provide useful hints, but they report point
locations without calibrated confidence, leaving researchers unable to tell
whether a geographic path is trustworthy. We introduce \fullname{}
(\name{}), a passive framework that evaluates router geolocation as a
path-level consistency problem. \name{} models each traceroute as a sequence
of candidate city-level locations and uses a Hidden Markov Model to fuse
local evidence with speed-of-light constraints and empirical latency priors.

\name{} produces a path consistency score summarizing how well
metadata and observed RTT increments support a coherent geographic
interpretation. Because this score is only meaningful when latency proxies
for geography, we also define a Path--Model Alignment metric that compares
speed-of-light residual increments of the decoded path against a reference
path. We evaluate on 413,354 RIPE Atlas traceroutes and a 6,555-path
subset verified by active probing. On validated paths, 94.2\% of decoded
sequences achieve mean error below 200\,km. PCS is largely
GeoDB-agnostic---median scores vary by less than 5\% across four commercial
databases---while the alignment metric reveals that over half of DB-IP and
IP2Location paths require substantial correction, compared with 15\% for
IPinfo. This lets downstream analyses quantify confidence in their
geographic conclusions rather than inheriting database accuracy without
qualification.
\end{abstract}

\maketitle

\section{Introduction}
\label{sec:introduction}

Traceroute-based Internet measurement depends not only on where routers are
placed, but also on whether those inferred locations are trustworthy. Traceroute
supports operational and research tasks such as diagnosing routing anomalies,
mapping physical infrastructure, and identifying latency bottlenecks. These
tasks often require geographic interpretations of intermediate routers, yet
infrastructure geolocation is substantially less reliable than end-host
geolocation because the available evidence is sparse, heterogeneous, and
unevenly curated. Commercial geolocation databases (GeoDBs), reverse-DNS
(rDNS) labels, Geofeed records, and Internet exchange point (IXP) metadata can
all provide useful location hints, but they usually provide point estimates
rather than calibrated confidence. As a result, downstream analyses often
inherit router locations without knowing whether a location is well supported,
weakly supported, or inconsistent with the rest of the measured path.

Existing geolocation workflows combine evidence primarily through source
preference, not through path-level consistency. In practice, researchers often
prefer high-precision sources such as Geofeeds when they are available, fall
back to parseable rDNS hints when labels appear informative, and rely on GeoDBs
for broad coverage. This hierarchy is pragmatic, but it still treats each hop
as a mostly local decision. A Geofeed entry can be stale, an rDNS label can be
ambiguous or mistyped, and a GeoDB entry can reflect registration or
headquarters information rather than router placement. The core limitation is
therefore not simply that any one source can be wrong; it is that the selected
sequence of hop locations is rarely checked for joint plausibility against the
traceroute itself.

Traceroute paths contain physical and contextual constraints that can expose
implausible router locations. Adjacent hops are not independent: their inferred
locations must be compatible with observed round-trip-time (RTT) changes,
speed-of-light limits in fiber, and the geographic structure implied by
neighboring hops. Violations appear as long-distance jumps under small RTT
increments, geographic oscillations, or locally plausible metadata that
contradicts stronger path-level evidence. These conflicts are especially
important for infrastructure addresses with weak direct evidence, including
private, unassigned, or poorly annotated hops. In such cases, a router location
should not be treated as a fact merely because one evidence source supplies a
coordinate; it should be treated as a hypothesis whose plausibility depends on
the full path context.

Active measurement can add location evidence, but it does not provide a general
confidence model for existing traceroute datasets. Triangulation and related
approaches use geographically distributed vantage points to constrain IP
locations~\cite{katz2006towards,dong2012ip}. However, active probing introduces
network overhead, can be affected by router rate limiting and filtering, and
requires access to measurement infrastructure such as RIPE Atlas or CAIDA Ark.
These requirements make active campaigns difficult to apply uniformly to large
historical datasets and to regions with sparse probe coverage. More
fundamentally, even when active or curated evidence is available, researchers
still need a principled way to ask whether the resulting path geography is
internally consistent. The same need---replacing point estimates with
continuous reliability signals---has driven designs in other domains, from
phi-accrual failure detectors that replace binary up/down verdicts with a
continuous suspicion level~\cite{hayashibara2004phi} to TrueTime's
uncertainty intervals that replace point
timestamps~\cite{corbett2013spanner}.

We introduce \fullname{} (\name{}), a passive framework that applies the
same principle to traceroute geolocation. \name{} frames
router geolocation as a global path alignment problem: rather than selecting one
source per hop and treating the result as ground truth, it asks which sequence
of candidate locations best explains both the available metadata and the
observed path behavior. We formalize this problem as a Hidden Markov Model
(HMM), where hidden states represent city-level location hypotheses, emissions
represent evidence from GeoDBs, rDNS, Geofeeds, and IXP metadata, and
transitions encode physical and empirical latency constraints. The decoded path
can help diagnose likely router locations, but the central output is a
likelihood-based path consistency score that quantifies how well the available
evidence supports a coherent geographic interpretation of the traceroute.

This paper makes the following contributions:

\vspace{0.3em}
\noindent
\textbf{Reliability scoring (\Cref{sec:methodology:scoring,sec:evaluation}).} \name{} produces a
likelihood-based score for each traceroute, allowing analyses to
separate reliable geographic inferences from paths whose locations
should be treated with caution.

\vspace{0.3em}
\noindent
\textbf{Multi-source fusion (\Cref{sec:methodology:emissions}).} \name{} combines
GeoDBs, rDNS hints, Geofeeds, and IXP metadata in a single HMM rather
than hard-coding a fixed source hierarchy. Consistent evidence can
reinforce a location hypothesis, while path context can expose
evidence that is locally plausible but globally inconsistent.

\vspace{0.3em}
\noindent
\textbf{Weak evidence (\Cref{sec:methodology:states,sec:methodology:emissions}).} By using
neighboring hops and latency constraints, \name{} can reason about
intermediate hops with missing, private, unassigned, or ambiguous IP
evidence without presenting the inferred location as certain.

\vspace{0.3em}
\noindent
\textbf{Path--model alignment (\Cref{sec:alignment,sec:validation:alignment_metric}).} We define a metric that
measures whether the decoded path and a reference path have
consistent speed-of-light residual increments.
This metric tells researchers when more negative PCS values reflect
genuine path-model disagreement and when PCS scores should be
interpreted with caution because latency is not acting as a reliable
proxy for geographic distance.

\vspace{0.3em}
\noindent
\textbf{Passive path diagnostics (\Cref{sec:validation,sec:evaluation}).} \name{}
flags segments where metadata evidence and physical latency
constraints diverge, surfacing cases that may reflect outdated
geolocation records, ambiguous labels, or path-level latency behavior
that makes geography opaque.

We evaluate \name{} on 413,354 RIPE Atlas traceroutes and a 6,555-path
validated subset verified by active probing.
On validated paths, 94.2\% of decoded sequences achieve mean error below
200\,km. Our validation asks whether path consistency scores correlate with
geolocation reliability and whether segments with more negative PCS reveal
cases where metadata evidence and physical path constraints disagree. We
compare raw GeoDB assignments against HMM-decoded paths across four
commercial databases and use segment-level analyses to examine how
path-level consistency changes the interpretation of router geolocation
evidence.

\section{Background}
\label{sec:background}

\subsection{Router Geolocation and Limitations}
\label{sec:background:geolocation}

Router geolocation methods combine heterogeneous evidence sources, but much of
this evidence is local to an individual IP address. Commercial geolocation
databases provide broad coverage, yet their accuracy for network infrastructure
is substantially weaker than for end-host
applications~\cite{pose:geodb,gharaibeh2017routergeo,imc-2023-geolocation-replication}.
Active measurement methods constrain locations with latency and topology
measurements~\cite{katz2006towards,dong2012ip}, while hint-based systems
derive location candidates from router hostnames and reverse-DNS
labels~\cite{luckie:hoiho,scheitle:hloc,du2020ripe}. These approaches improve
per-hop evidence quality, but downstream workflows still choose a location
through a source hierarchy or a per-address confidence score, without checking
whether the selected sequence of locations is jointly plausible against the
traceroute.

\subsection{Constraints and Artifacts in Traceroute Measurements}
\label{sec:background:traceroute-constraints}

\begin{figure}[t]
    \centering
    \includegraphics[width=\linewidth]{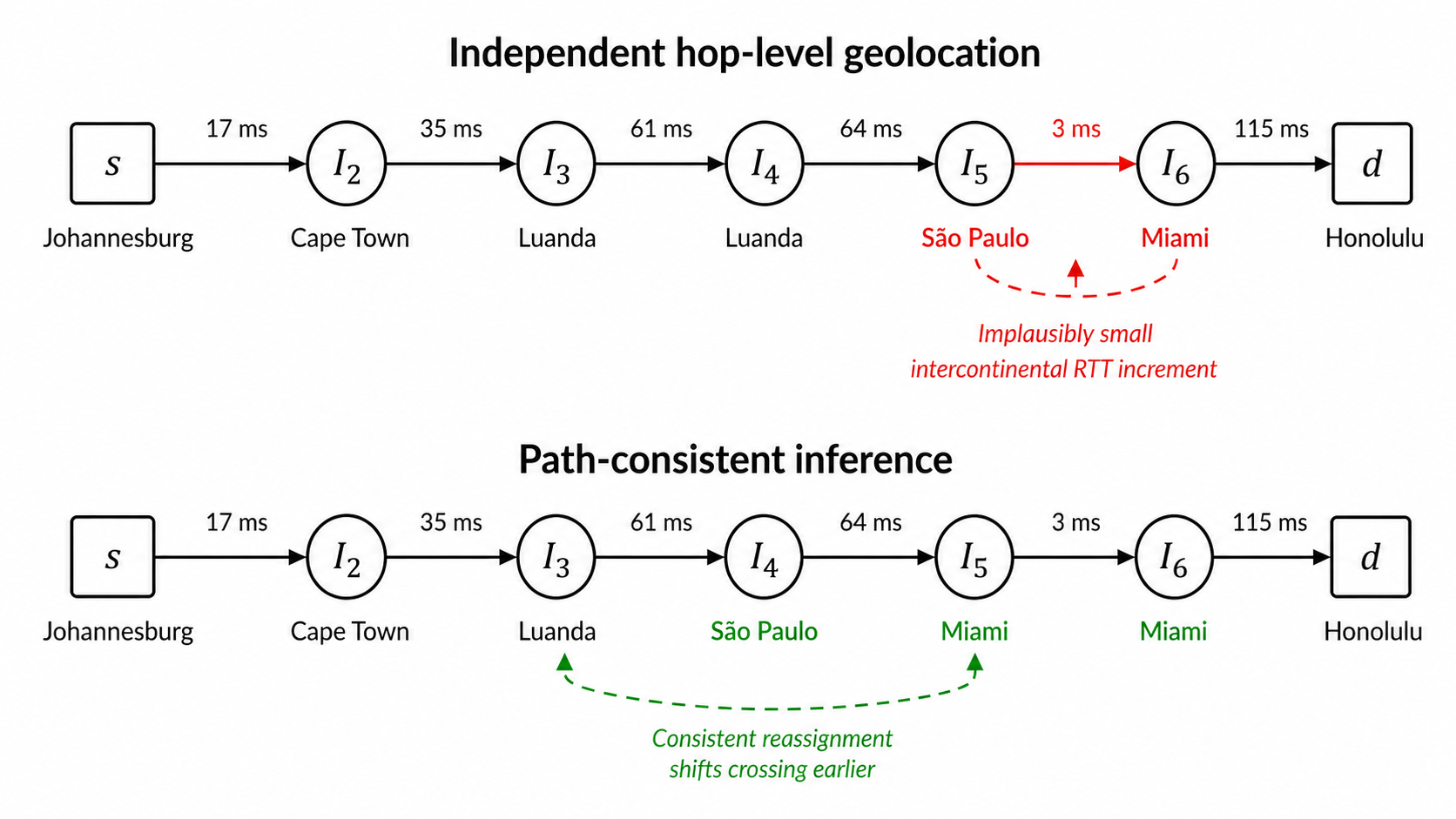}
    \caption{
        (a) An implausible transition between adjacent hops geolocated by commercial databases in Sao Paulo and Miami, and
        (b) a geographically coherent aligned path informed by RTT observations.
    }
    \label{fig:example}
\end{figure}

Traceroute provides sequential constraints that independent geolocation
lookups ignore: adjacent location assignments must be compatible with RTT
increments, speed-of-light limits in fiber, and the geographic context of
neighboring hops.
Figure~\ref{fig:example} illustrates this with a traceroute from South Africa ($s$) to Hawaii ($d$).
The raw database output maps hop $I_5$ to Sao Paulo and $I_6$ to Miami, but the 3\,ms inter-hop latency is too small for a transcontinental transition; the 64\,ms latency between $I_4$ in Luanda and $I_5$ instead suggests a trans-Atlantic crossing.
The question is not which individual hop location is most likely, but whether the complete path forms a physically coherent sequence.

Traceroute artifacts can also make geography opaque rather than simply
incorrect. Delay anomalies distort the RTT--distance
relationship~\cite{imc17:romain}, and MPLS tunnels can hide internal hops
or create IP-level adjacencies that do not correspond to physical router
adjacency~\cite{mpls-paul-2011-imc,matthew-mpls-2012-ccr,wormhole-2017-imc}.
\name{} therefore treats traceroute order and RTT increments as soft
reliability constraints rather than a complete physical map.

\subsection{Sequential Inference for Path-Level Geolocation}
\label{sec:background:hmm}

Hidden Markov Models provide a natural abstraction for tasks where the
observed data are noisy and the desired explanation is a latent
sequence~\cite{rabiner1986,bilmes1998,ephraim2002}. For router geolocation,
hidden states are candidate city-level locations, emissions represent
metadata support, and transitions encode the feasibility of moving between
adjacent locations under the observed RTT increment. Unlike a per-hop
lookup, an HMM can balance local metadata against path-level constraints:
strong evidence at one hop can anchor nearby ambiguous hops, while a
physically infeasible transition can lower confidence in an otherwise
plausible point estimate.

\section{Datasets}
\label{sec:datasets}

\name{} uses measurement data to define the path that must be explained and
metadata sources to define the candidate locations that may explain it. 
The measurement data provide ordered traceroute hops, responding IP addresses, 
and RTTs. The metadata sources provide city-level location 
hypotheses for each hop. We keep these roles separate: metadata 
sources are evidence for candidate locations, while the HMM in 
\Cref{sec:methodology} decides whether a sequence of candidates is 
jointly plausible under the measured path.

\subsection{RIPE Atlas Traceroute Corpus}
\label{sec:datasets:traceroutes}

Our traceroute corpus comes from RIPE Atlas anchoring measurements
\cite{ripe2017anchoring}, which repeatedly measure paths among anchors with
known locations. These recurring measurements are useful for two distinct
purposes. First, their high volume provides enough inter-region latency
samples to learn empirical transition priors. Second, their stable
vantage points support a held-out evaluation set whose endpoints can be anchored
during path alignment. To avoid leakage between these roles, we separate
the prior construction and evaluation windows by date.

\begin{table*}[t]
    \centering
    \caption{
        RIPE Atlas measurement datasets used by \name{}.
        The prior-construction window is used for learning empirical
        latency priors, while the evaluation corpus is used for
        decoding and validation.
    }
    \label{table:datasets}
    \small
    \begin{tabular}{p{0.22\textwidth}p{0.28\textwidth}p{0.42\textwidth}}
        \toprule
        \textbf{Dataset} & \textbf{Date and volume} & \textbf{Role in the paper} \\
        \midrule
        Prior construction
            & 2025-07-01, 24 hours; 38,610,738 traceroutes
            & Learn empirical country-pair latency priors. \\
        Evaluation Corpus
            & 2025-08-01, 6 hours; 413,354 reached-destination traceroutes
            & Large-scale inference, PCS analysis, and path-alignment diagnostics. \\
        Validated subset
            & 6,555 traceroutes
            & Ground truth paths verified via active probing. \\
        \bottomrule
    \end{tabular}
\end{table*}

The prior-construction dataset, detailed in Table~\ref{table:datasets}, is used to
estimate empirical latency distributions for ordered country pairs. 
The evaluation corpus consists of a held-out six-hour window from which we
retain one traceroute per distinct source-destination anchor pair after
filtering traceroutes that did not reach the destination anchor. This filter
reduces 693,849 total traceroutes in the raw window to 413,354
reached-destination traceroutes in the Evaluation Corpus; among these, 410,643
yield decoded paths for each GeoDB vendor and therefore form the
denominator for the full-corpus CDFs in \Cref{sec:evaluation}.

We extract a validated subset from the evaluation corpus through an active
measurement campaign. This subset represents approximately 
1.6\% of the evaluation corpus. Within this set, we identify a
\textbf{Public-Candidate} subset of 4,747 paths for which all validated
hop locations are present within candidates derived strictly from
public metadata. Our primary accuracy evaluations and database
comparisons utilize the full validated set.

\subsection{Location Evidence Sources}
\label{sec:datasets:evidence}

Commercial geolocation databases provide broad-coverage point estimates for
observed hops. We query IPinfo \cite{IPinfo2025}, MaxMind
\cite{maxmindgeoloc}, DB-IP \cite{dbipgeoloc}, and IP2Location \cite{ip2locationgeoloc} for city, country, and coordinate
fields. To ensure temporal consistency, we utilize snapshots from
August 2025 for all vendors. We evaluate each database independently by
executing four distinct decoding runs per traceroute; in each iteration,
the candidate pool is populated by a single GeoDB vendor alongside the
auxiliary metadata and peering signals described below.

Table~\ref{tab:datasets:geodb-coverage} shows that, out of 52,784 unique
responding IPs (100\%), DB-IP, IPinfo, and IP2Location map more than 92\%,
whereas MaxMind maps only 45.86\%; its decoded paths are therefore conditioned
on fewer vendor-provided candidates.

\begin{table}[t]
    \centering
    \caption{GeoDB mapping coverage over unique responding hop IPs.}
    \label{tab:datasets:geodb-coverage}
    \small
    \begin{tabular}{lr}
        \toprule
        \textbf{GeoDB} & \textbf{Share} \\
        \midrule
        DB-IP & 92.85\% \\
        IPinfo & 92.85\% \\
        IP2Location & 92.31\% \\
        MaxMind & 45.86\% \\
        \bottomrule
    \end{tabular}
\end{table}

Auxiliary sources add context missing from commercial database records.
Geofeed records associate prefixes with operator-published location
information \cite{geolocatemuch}. PeeringDB IXP candidates use PeeringDB
\cite{peeringdb} in two ways: we match hop IPs against known IXP prefixes to
identify exchange-point interfaces, and, for adjacent hops that form an
inter-AS link, we query facilities where both ASes peer and add those facility
locations as geographic candidates. CAIDA/IPinfo bogon flags identify private 
and reserved address space
\cite{CAIDA_Bogon2025,IPinfoBogon2025}. Finally, OpenINTEL/Aleph rDNS hints
parse reverse-DNS records with Aleph \cite{aleph:conext}, which decodes
operator-specific naming conventions such as embedded city or airport codes.
These hints are treated as additional candidate evidence, not as deterministic
labels.

\Cref{tab:datasets:public-source-coverage} shows sparse but useful public
evidence: OpenINTEL/Aleph rDNS hints cover 18.56\% of interfaces, PeeringDB IXP
candidates 8.26\%, CAIDA/IPinfo bogon flags 7.15\%, and Geofeed records only
0.91\%.

\begin{table}[t]
    \centering
    \caption{Public evidence availability over unique responding hop IPs,
    sorted by share.}
    \label{tab:datasets:public-source-coverage}
    \small
    \begin{tabular}{lr}
        \toprule
        \textbf{Public evidence source} & \textbf{Share} \\
        \midrule
        OpenINTEL/Aleph rDNS hints & 18.56\% \\
        PeeringDB IXP candidates & 8.26\% \\
        CAIDA/IPinfo bogon flags & 7.15\% \\
        Geofeed records & 0.91\% \\
        \bottomrule
    \end{tabular}
\end{table}

\subsection{Evidence Normalization}
\label{sec:datasets:normalization}

We normalize all evidence sources into a common candidate representation
before running \name{}. Each candidate stores a city, country,
latitude, longitude, source label, and source-specific confidence weight. 
Candidates that name the same city are merged so that agreement across
sources increases emission support. Conflicting candidates remain
available for the HMM to evaluate against the rest of the path.

The normalized path representation preserves weak or incomplete evidence. 
For each hop, \name{} keeps the responding IP address, source-specific
candidate locations, and metadata flags.
For adjacent responding hops, we compute the inter-hop RTT increment used
by the transition model. Missing, private-address, and
non-responding hops become low-certainty positions in the sequence
rather than holes that break the path.

\section{Path Alignment as a Hidden Markov Model}
\label{sec:methodology}

\name{} formalizes router geolocation reliability as a path-level alignment
problem. Given a traceroute with $T$ observed hops, let $o_{1:T}$ denote the
observation sequence, where each observation $o_t$ contains the hop IP address
when available, its normalized location evidence, and the measured round-trip
time $RTT_t$. The latent sequence $s_{1:T}$ represents the geographic
interpretation of the path, where each state $s_t$ is a city-level location
hypothesis with latitude and longitude. \name{} seeks the sequence that best
explains both the per-hop evidence and the physical constraints between
adjacent hops:
\begin{equation}
    s_{1:T}^* = \operatorname*{arg\,max}_{s_{1:T}} P(s_{1:T}, o_{1:T}).
\end{equation}

The HMM consists of four modules. First, \name{} constructs a finite candidate
state space for each hop from the evidence sources described in
\Cref{sec:datasets}. Second, an emission model converts agreement and conflict
among those sources into probabilistic support for each candidate location.
Third, a transition model scores whether adjacent candidate locations are
compatible with the observed RTT increment and empirical latency behavior.
Finally, Viterbi decoding produces the most likely geographic path, and the
normalized path likelihood becomes the path consistency score.

\begin{figure*}[t]
    \centering
    \includegraphics[width=\textwidth]{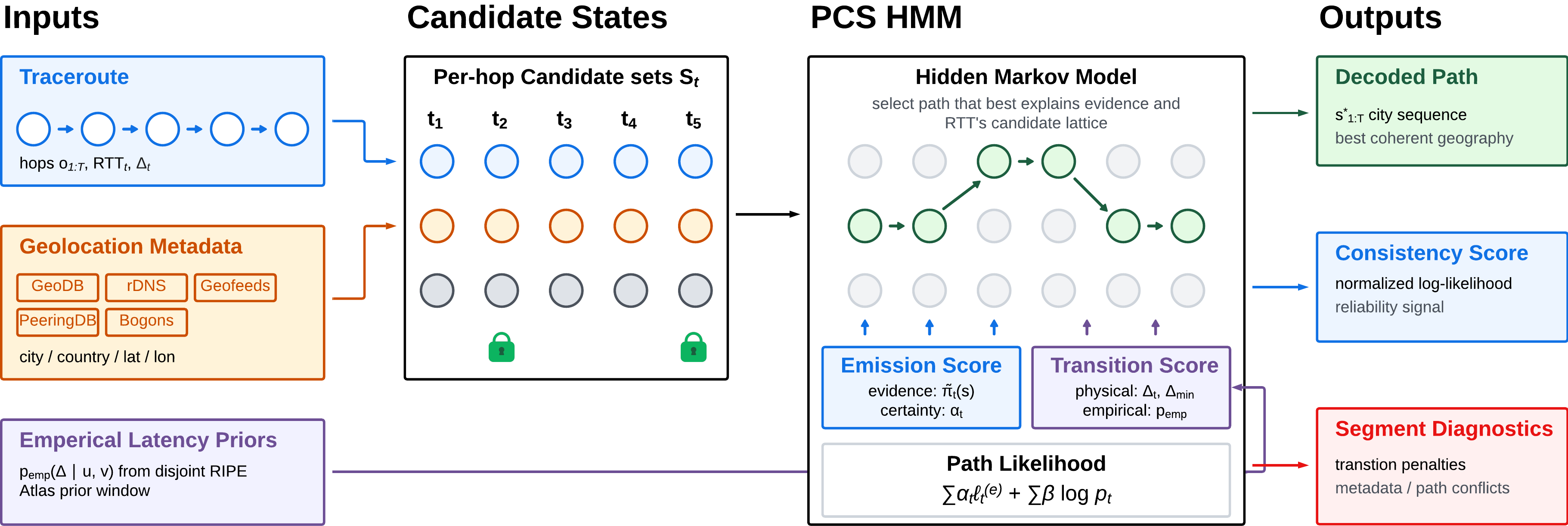}
    \caption{
        Overview of the \name{} path-alignment pipeline.
        Traceroute observations and normalized location evidence define a
        candidate state space, emission scores, and latency-aware transition
        scores. Viterbi decoding then produces a path-level geographic
        interpretation and a path consistency score.
    }
    \label{fig:pcs-pipeline}
\end{figure*}

\subsection{Candidate State Space and Endpoint Anchors}
\label{sec:methodology:states}

\name{} restricts inference to a discrete set of candidate locations at each
hop so that path alignment remains tractable and evidence-driven. For hop $t$,
we define a candidate set $\mathcal{S}_t$ by unioning the normalized
city-level locations supplied by commercial GeoDBs, rDNS-derived hints,
Geofeed records, and IXP or facility metadata. If multiple sources point to
the same city-level location, \name{} keeps a single candidate and records the
set of supporting sources. If sources disagree, their locations remain as
competing candidates, allowing the HMM to decide which candidate is most
consistent with the rest of the path.

Endpoint anchoring turns unconstrained geolocation into a constrained path
interpolation problem. In our setting, the source and destination endpoints can
often be associated with stronger metadata than intermediate routers, such as
RIPE Atlas probe metadata for the source and known destination metadata for
controlled measurements. When such anchors are available, \name{} fixes
$s_1$ and $s_T$ to those endpoint locations and infers the intermediate
sequence between them. This design prevents weak intermediate evidence from
moving the entire path to an implausible geography, while still allowing the
model to express uncertainty for internal hops.

\subsection{Emission Model: Fusing Location Evidence}
\label{sec:methodology:emissions}

The emission model measures how strongly the available metadata supports each
candidate location at a hop. For candidate $s \in \mathcal{S}_t$, \name{}
computes an additive evidence utility
\begin{equation}
    u_t(s) = \sum_k w_k \phi_k(t,s),
\end{equation}
where $k$ indexes evidence sources, $w_k$ is the source weight, and
$\phi_k(t,s)$ indicates the support that source $k$ provides for assigning hop
$t$ to location $s$. This representation lets independent sources reinforce a
shared candidate while preserving disagreement as explicit competition among
states.

\name{} converts utilities into an emission distribution via a
temperature-scaled softmax over the $K_t = |\mathcal{S}_t|$ candidates:
\begin{equation}
    \pi_t(s) =
    \frac{\exp(u_t(s) / \tau_t)}
    {\sum_{s' \in \mathcal{S}_t} \exp(u_t(s') / \tau_t)},
\end{equation}
where $\tau_t$ controls how sharply the model separates high- and low-utility
candidates. A uniform smoothing floor ensures no candidate receives zero
probability:
\begin{equation}
    \tilde{\pi}_t(s) = (1-\gamma)\pi_t(s) + \gamma \frac{1}{K_t}.
\end{equation}

The emission contribution is weighted by hop-specific evidence certainty.
\name{} derives an emission weight $\alpha_t \in [0,1]$ from local evidence
statistics such as agreement, entropy, and source saturation. Hops with
concentrated, mutually reinforcing evidence receive higher $\alpha_t$, so the
decoded path is encouraged to respect the metadata. Hops with missing,
ambiguous, or conflicting evidence receive lower $\alpha_t$, allowing the
transition model and neighboring hops to carry more of the inference.

\subsection{Transition Model: Latency-Aware Path Constraints}
\label{sec:methodology:transitions}

The transition model evaluates whether moving from one candidate location to
the next is compatible with the measured traceroute path. For adjacent hops
$t-1$ and $t$, we define the nonnegative RTT increment
\begin{equation}
    \Delta_t = \max(0, RTT_t - RTT_{t-1}).
\end{equation}
Given two candidate locations $s_{t-1}$ and $s_t$, let $d(s_{t-1},s_t)$ be
their geographic distance. The minimum round-trip propagation delay required
for that distance is
\begin{equation}
    \Delta_{\min}(s_{t-1},s_t) =
    \frac{2d(s_{t-1},s_t)}{\rho c},
\end{equation}
where $c$ is the speed of light in vacuum and $\rho c$ is the effective
propagation speed in fiber.

\name{} combines a physical feasibility term with empirical latency evidence.
The physical term penalizes transitions whose distance cannot be supported by
the observed RTT increment. We define the residual slack as
\begin{equation}
    r_t = \max(\epsilon, \Delta_t - \Delta_{\min}),
\end{equation}
and model feasible slack with a Pareto-II survival term:
\begin{equation}
    p_{\text{slack}} =
    \left(\frac{m}{r_t + m}\right)^\eta.
\end{equation}
Transitions that violate the speed-of-light constraint are suppressed with a
logistic gate,
\begin{equation}
    g_t = \sigma\left(k(\Delta_t - \Delta_{\min})\right),
\end{equation}
yielding $p_{\text{phys}} = g_t p_{\text{slack}}$. This form does not require
a hard cutoff at the feasibility boundary, which is important because
traceroute RTTs can include queueing, measurement noise, and router response
artifacts.

\begin{figure}[t]
    \centering
    \begin{subfigure}[t]{0.48\linewidth}
        \centering
        \includegraphics[width=\linewidth]{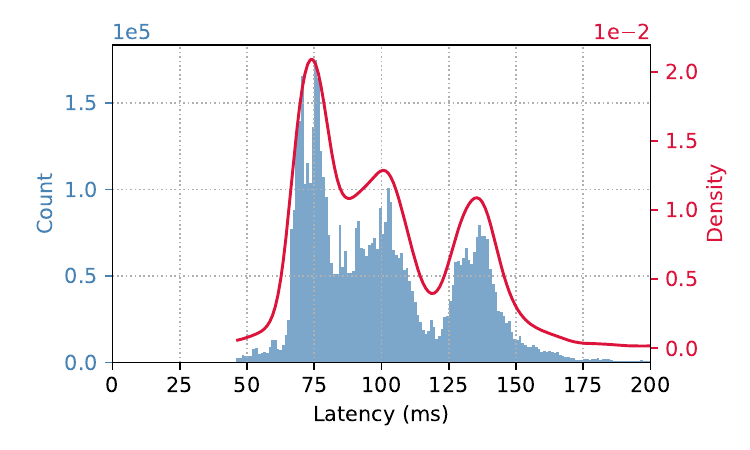}
        \caption{US $\leftrightarrow$ GB}
        \label{fig:kde:gb-us}
    \end{subfigure}
    \vspace{0.5em}
    \begin{subfigure}[t]{0.48\linewidth}
        \centering
        \includegraphics[width=\linewidth]{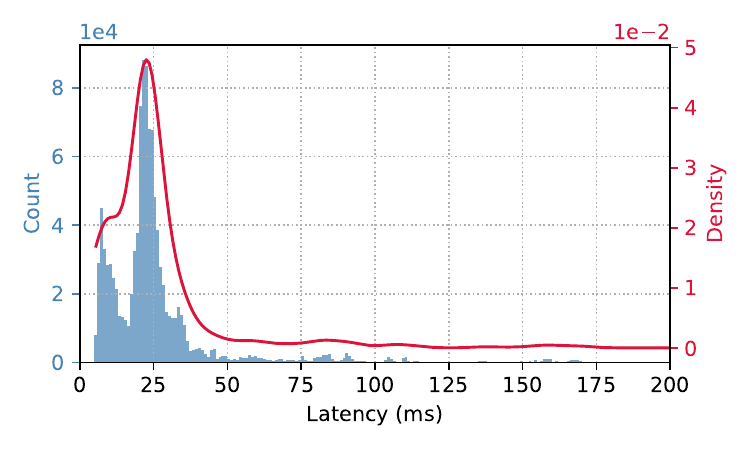}
        \caption{NL $\leftrightarrow$ SE}
        \label{fig:kde:nl-se}
    \end{subfigure}
    \caption{
        Empirical latency distributions used to construct transition priors.
        The US--GB pair has distinct modes near 68\,ms, 105\,ms, and 133\,ms,
        while the NL--SE pair concentrates near 38\,ms with a smaller mode
        near 14\,ms. These multimodal profiles motivate a transition model
        that blends physical feasibility with country-pair empirical evidence.
    }
    \label{fig:kde-examples}
\end{figure}

Empirical latency evidence captures country-pair behavior that a pure
propagation model cannot express. We estimate these priors from the disjoint
prior-construction dataset in \Cref{table:datasets}. For each ordered country
pair $(u,v)$ with sufficient observations, we collect the inter-hop RTT
increments whose endpoint candidates map to $(u,v)$, bin the increments at
5\,ms granularity, and apply Gaussian kernel smoothing with bandwidth
$\sigma=2$. This produces a continuous empirical density
$p_{\text{emp}}(\Delta \mid u,v)$ that preserves latency modes induced by
multiple physical routes, routing policies, and measurement vantage points.
\Cref{fig:kde-examples} illustrates why this empirical term is necessary: even
common country pairs can have several plausible latency regimes, so a single
distance-based penalty would either over-penalize valid alternatives or become
too permissive to be useful.

The final transition probability is a trust-weighted blend:
\begin{equation}
    p_t(s_{t-1} \to s_t) =
    \lambda_t p_{\text{emp}} + (1-\lambda_t)p_{\text{phys}}.
\end{equation}
The trust weight $\lambda_t$ increases with the amount of relevant empirical
data but is capped below one, ensuring that every transition retains a
physical feasibility component. \name{} also includes a co-location bonus for
successive hops assigned to the same city and a revisit penalty for paths that
leave a city and immediately return, reducing implausible geographic
oscillation.

\subsection{Viterbi Decoding and Path Consistency Scoring}
\label{sec:methodology:scoring}

\name{} decodes the most likely geographic sequence with the Viterbi
algorithm. Let $\ell_t^{(e)}(s)=\log \tilde{\pi}_t(s)$ be the emission
log-probability for assigning hop $t$ to candidate $s$. The dynamic program
updates
\begin{equation}
    \delta_t(s) =
    \max_{s' \in \mathcal{S}_{t-1}}
    \left[
        \delta_{t-1}(s')
        + \alpha_t \ell_t^{(e)}(s)
        + \beta \log p_t(s' \to s)
        \right],
\end{equation}
where $\beta$ controls the stiffness of the transition model relative to the
emission evidence. Backtracking through the maximizing predecessors yields the
decoded path $s_{1:T}^*$.

The primary output is the path consistency score, a normalized
log-score for path-level geolocation consistency. We compute PCS as the
length-normalized log-likelihood of the decoded sequence:
\begin{equation}
    PCS(o_{1:T}) = \frac{1}{T}
    \left[
        \sum_{t=1}^{T} \alpha_t \ell_t^{(e)}(s_t^*)
        + \sum_{t=2}^{T} \beta \log p_t(s_{t-1}^* \to s_t^*)
        \right].
\end{equation}
Because PCS is a log-likelihood, its values are typically negative:
values closer to zero indicate that the available metadata and measured
RTT increments better support a coherent geographic interpretation of
the traceroute. More negative PCS values flag paths or segments where
the evidence is internally inconsistent, which may reflect stale
geolocation records, ambiguous rDNS labels, or path-level measurement
effects.

\subsection{Path--Model Alignment}
\label{sec:alignment}

A more negative PCS can arise because a database location is wrong,
but it can also arise because traceroute latency is a poor proxy for
geographic distance on that path. We therefore report an alignment
diagnostic alongside PCS to separate these failure modes.

We define a \textbf{Path--Model Alignment} metric $\mathcal{R}$
that compares the speed-of-light residual profile of the decoded
HMM path with that of a reference path.
The reference is the raw GeoDB path $s^{(g)}_{1:T}$ in vendor-specific
runs, or the independently validated path $s^{(v)}_{1:T}$ for the
Public-Candidate subset where no GeoDB candidate is supplied.
$\mathcal{R}$ asks whether the HMM path and the reference induce
similar latency--distance behavior, not whether either is correct in
isolation.

For any geographic path $x_{1:T}$, we first compute the propagation
time implied by consecutive locations:
\begin{equation}
    \hat{\Delta}_{t}(x) =
    \frac{2\, d(x_{t-1},\, x_t)}{v_{\mathcal{R}}},
\end{equation}
where $v_{\mathcal{R}}$ is the effective propagation speed used by the
alignment diagnostic. We set $v_{\mathcal{R}}=200\,\mathrm{km/ms}$,
a round-number approximation of the transition speed
$\rho c=197.86\,\mathrm{km/ms}$ introduced in
\Cref{sec:methodology:transitions}. The ${\sim}1\%$ difference is
negligible relative to other sources of residual variation.
We then convert this propagation time into a normalized
speed-of-light residual,
\begin{equation}
    e_t(x) =
    \frac{
        \left|\Delta_t - \hat{\Delta}_{t}(x)\right|
    }{
        \hat{\Delta}_{t}(x) + \tau
    },
\end{equation}
where $\tau$ is a soft floor that prevents near-zero physical
increments from dominating the ratio.

\name{} compares paths using the \emph{increments} of these residuals
rather than their absolute levels, making the metric robust to
path-wide offsets.
Let $\nabla e_t^{(h)} = e_t(s^*) - e_{t-1}(s^*)$ and
$\nabla e_t^{(q)} = e_t(q) - e_{t-1}(q)$ be the residual increments
for the decoded and reference paths, respectively.
We define
\begin{equation}
    \label{eq:alignment}
    \mathcal{R} =
    \max\left(
        0,\;
        1 -
        \frac{
            \sum_t \left|\nabla e_t^{(h)} - \nabla e_t^{(q)}\right|
        }{
            \epsilon +
            \sum_t \max\!\left(\left|\nabla e_t^{(h)}\right|,
                               \left|\nabla e_t^{(q)}\right|\right)
        }
    \right).
\end{equation}
Only hop pairs with RTTs and coordinates in both paths contribute to
the sums.
The resulting score lies in $[0,1]$: $\mathcal{R}=1$ means the
decoded and reference paths have matching residual increments, while
$\mathcal{R}\approx 0$ means they disagree about where
latency--distance inconsistencies appear.

When $\mathcal{R}$ is high, PCS can be read as a calibrated confidence
signal; when $\mathcal{R}$ is near zero, the path likely contains
latency or metadata conflicts that make residuals unstable, and PCS
should be treated as a warning about model applicability.
$\mathcal{R}$ does not prove geographic correctness---it cannot detect
cases where both paths are consistently wrong, \eg systematic database
errors that happen to be latency-consistent.
At the dataset level, the distribution of $\mathcal{R}$ characterizes
how much a researcher can trust PCS-based conclusions from a given
GeoDB: predominantly high $\mathcal{R}$ means PCS is interpretable,
while substantial mass near zero signals that many paths fall outside
the model scope.

The parameter configuration used for all decoding runs is reported in
\Cref{sec:appendix:parameters}.

\section{Validation}
\label{sec:validation}

We evaluate \name{} using an active-ping validation set designed to test 
path-level consistency without assuming the correctness of commercial 
databases. As detailed in Section~\ref{sec:datasets}, this set consists of 
6,555 fully validated paths where every intermediate hop has been 
independently verified through a shortest-ping constraint. 

To identify these hops, we first derive potential regions from 
hop-level metadata and neighboring-hop context. We select at least ten 
RIPE Atlas probes near these regions and issue active pings to the 
hop IP address. A hop is marked as verified if at least one probe 
observes an RTT below 2\,ms. This threshold physically constrains 
the router to a local region within approximately 200\,km of the 
responding probe, providing a location constraint strong enough to 
evaluate the plausibility of decoded city-level paths.

\subsection{Algorithmic Alignment Accuracy}
\label{sec:validation:alignment}

We first assess whether the HMM correctly aligns the traceroute path 
when the ground-truth locations are present in the candidate state 
space. We utilize the \textbf{Public-Candidate} subset, 
consisting of 4,747 traceroutes where the validated location for 
every hop is recoverable strictly through public metadata sources.

\begin{figure}[t]
    \centering
    \includegraphics[width=0.86\linewidth]{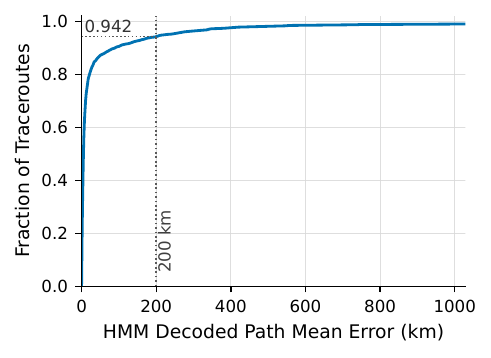}
    \caption{
        Path-level alignment accuracy for the Public-Candidate subset ($N=4,747$).
        The CDF shows the distribution of mean path error relative to 
        verified locations, with a dashed vertical line marking the 200\,km 
        validation threshold. The plot is capped at the 99th percentile 
        (1,035.86\,km); 94.2\% of traceroutes exhibit a mean path error 
        within the 200\,km threshold.
    }
    \label{fig:alignment-accuracy-cdf}
\end{figure}

Figure~\ref{fig:alignment-accuracy-cdf} presents the CDF of the mean path error 
for these traceroutes. For 94.2\% of the paths, the HMM-decoded path achieves 
a mean error below our 200\,km validation threshold. This confirms 
the framework's ability to identify the correct geographic sequence 
under ideal metadata coverage. The long tail in the distribution, 
representing approximately 1\% of paths above the 99th percentile cap, 
reflects cases where measured latency no longer tracks geographic
distance closely enough to support hop-level localization.

\subsection{Corrective Power and Database Comparison}
\label{sec:validation:geodb}

We evaluate the framework's ability to resolve geolocation errors 
across the entire validated set. We compare the raw geography 
provided by each commercial GeoDB against the \name{} decoded path 
generated using that same database vendor as a primary candidate source.

\begin{figure*}[t]
    \centering
    \begin{subfigure}[t]{0.32\textwidth}
        \centering
        \includegraphics[width=\linewidth]{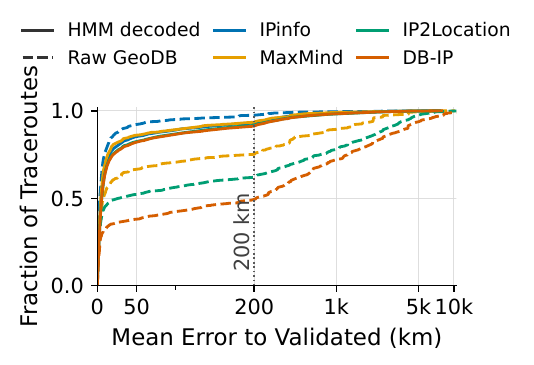}
        \caption{Traceroute mean error.}
        \label{fig:geodb-comparison-traceroute}
    \end{subfigure}
    \hfill
    \begin{subfigure}[t]{0.32\textwidth}
        \centering
        \includegraphics[width=\linewidth]{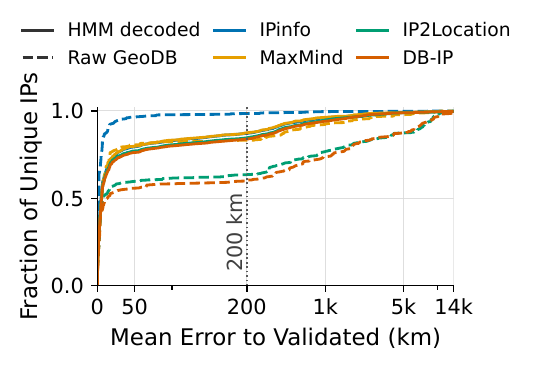}
        \caption{Unique-IP mean error.}
        \label{fig:geodb-comparison-ip}
    \end{subfigure}
    \hfill
    \begin{subfigure}[t]{0.32\textwidth}
        \centering
        \includegraphics[width=\linewidth]{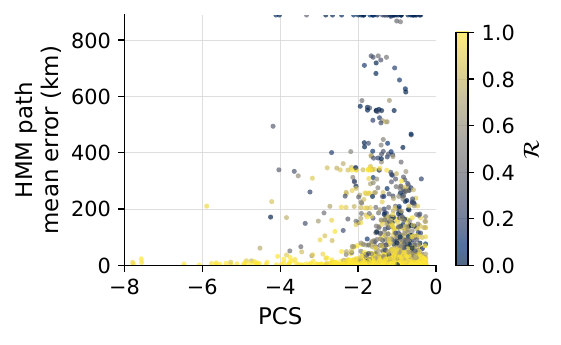}
        \caption{Public-candidate alignment.}
        \label{fig:r-scatter-public}
    \end{subfigure}
    \caption{
        Validation-set accuracy and alignment diagnostics. Panels (a) and (b)
        compare raw GeoDB assignments (dashed) and HMM-decoded paths (solid) at
        traceroute and unique-IP granularity. Their x-axes are linear through the
        200\,km validation threshold and logarithmic beyond it; values are capped
        at the 99th percentile (10,498.4\,km in (a), 13,980.4\,km in (b)).
        Panel (c) relates PCS, decoded error, and Path--Model Alignment
        $\mathcal{R}$ in the Public-Candidate subset.
    }
    \label{fig:combined-geodb-cdfs}
\end{figure*}

Figure~\ref{fig:geodb-comparison-traceroute} shows the error distributions for each vendor 
at the traceroute level. For noisier vendors, \name{} provides substantial 
corrective lift. For instance, the raw median error for DB-IP is 207.1\,km, 
with only 49.2\% of paths falling within the 200\,km validation threshold. 
After HMM decoding, the median error drops to 5.4\,km, and the fraction 
of paths within 200\,km increases to 91.2\%. We observe similar trends 
for IP2Location, where compliance at 200\,km rises from 61.9\% to 91.6\%, 
and MaxMind, which improves from 75.1\% to 93.5\%.

However, the gain is negligible or negative for high-quality sources 
such as IPinfo, which exhibits a raw path compliance of 97.5\% and a 
median error of 3.0\,km. In these instances, the HMM optimizes for path 
sequences with high physical and empirical transition likelihoods. 
When traceroute measurements are impacted by path-level latency
anomalies, these physically consistent alignments may deviate from
the verified ground-truth locations, resulting in a decoded compliance
of 93.0\%.

Figure~\ref{fig:geodb-comparison-ip} aggregates these results by unique IP to 
assess per-hop reliability after removing path-level redundancy. When 
aggregating by IP, the median error for decoded paths remains stable 
(ranging from 4.9\,km to 5.8\,km across vendors). For DB-IP, the unique 
IP compliance at 200\,km improves from 59.8\% in the raw database to 
83.8\% after decoding. These results indicate that \name{} is effective 
at pruning ``teleportation'' jumps and administratively-centered coordinates 
by leveraging the physical constraints of the surrounding traceroute.

\subsection{Alignment Metric Calibration}
\label{sec:validation:alignment_metric}

Finally, we use the Path--Model Alignment metric $\mathcal{R}$
defined in \Cref{sec:alignment} to diagnose when PCS is reliable
and when traceroute latency is not a useful proxy for geography.
PCS is a log-score and is typically negative, so values closer to zero
indicate stronger path consistency and more negative values indicate
weaker consistency.
Unlike the CDFs above, which evaluate endpoint error directly,
$\mathcal{R}$ compares the speed-of-light residual increments of
two path interpretations. In the Public-Candidate subset, the
comparison is between the HMM path and the independently validated
path, because no GeoDB candidate is used by the decoder. In the
vendor-specific setting, the comparison is between the HMM path and
the corresponding raw GeoDB path. This distinction lets us separate
algorithmic alignment failures from cases where the database path
and the latency-constrained HMM path disagree.

\Cref{fig:r-scatter-public} validates the metric in the cleanest
setting, where all candidate locations are available from public
metadata and the reference path is the active-ping validation path.
The distribution is concentrated in the high-alignment, low-error
region: 86.5\% of paths have $\mathcal{R}\ge0.8$, and these paths
have a median mean error of 4.0\,km. The low-alignment tail is small
but informative. Paths with $\mathcal{R}<0.2$ account for only 3.8\%
of the subset, yet their median mean error rises to 244.7\,km.
Likewise, among paths whose mean error exceeds the 200\,km validation
threshold, the median $\mathcal{R}$ drops to 0.317 and 37.7\% have
$\mathcal{R}<0.2$. Thus, even when the correct locations are present
in the candidate space, the remaining failures are concentrated in
paths where the hop-by-hop latency residual pattern no longer matches
the validated geography.

\begin{figure*}[t]
    \centering
    \includegraphics[width=\textwidth]{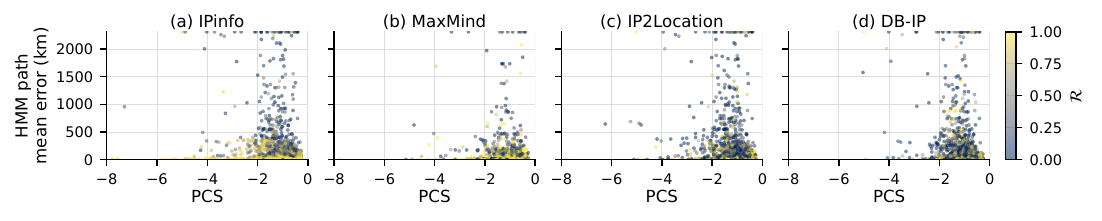}
    \caption{
        Validated vendor runs: PCS versus mean decoded geolocation
        error, colored by Path--Model Alignment $\mathcal{R}$ between
        the HMM path and the raw GeoDB path. Low-alignment paths
        concentrate where PCS is more negative and decoded error is high.
    }
    \label{fig:r-scatter-geodb-scatter}
\end{figure*}

\begin{figure*}[t]
    \centering
    \includegraphics[width=\textwidth]{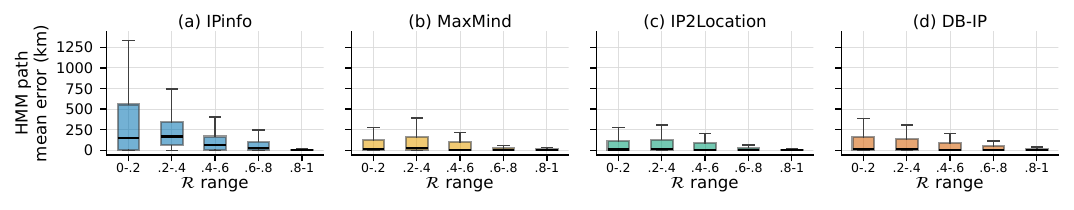}
    \caption{
        Validated vendor runs: mean decoded geolocation error grouped
        by Path--Model Alignment $\mathcal{R}$ range. The binned view
        separates the high-alignment paths, which remain near city-scale
        error, from the low-alignment paths with heavier error tails.
    }
    \label{fig:r-scatter-geodb-boxplot}
\end{figure*}

\Cref{fig:r-scatter-geodb-scatter,fig:r-scatter-geodb-boxplot} use the
raw GeoDB path as the reference for each vendor-specific run and report
both pointwise and binned views of the same alignment signal. The scatter plot
(\Cref{fig:r-scatter-geodb-scatter}) shows where low-$\mathcal{R}$
paths fall in the PCS--error plane; the boxplot
(\Cref{fig:r-scatter-geodb-boxplot}) asks whether those alignment
levels correspond to different error distributions. Across the four
vendor runs, the median $\mathcal{R}$ is 0.895, but 22.0\% of
path--vendor pairs have $\mathcal{R}<0.2$. This mass is not uniform:
IPinfo has the strongest alignment profile, with median
$\mathcal{R}=0.986$, 76.0\% of paths at $\mathcal{R}\ge0.8$, and only
7.1\% below $\mathcal{R}<0.2$. MaxMind is the next strongest vendor by
high-alignment mass, with median $\mathcal{R}=0.921$ and 54.9\% of paths
at $\mathcal{R}\ge0.8$, but its low-alignment tail is larger
(19.5\% below $\mathcal{R}<0.2$). IP2Location and DB-IP exhibit weaker
path-model agreement: their median $\mathcal{R}$ values are 0.526 and
0.396, respectively, and 34.7\% and 37.4\% of their paths fall below
$\mathcal{R}<0.2$.

The boxplot confirms that $\mathcal{R}$ separates error regimes rather
than only coloring the scatter plot. In the high-alignment bin
($\mathcal{R}\ge0.8$), median decoded error is near city scale for all
vendors: 3.6\,km for IPinfo, 5.7\,km for MaxMind, 3.4\,km for
IP2Location, and 4.7\,km for DB-IP. In contrast, the lowest-alignment
bin ($\mathcal{R}<0.2$) has much heavier error tails, with mean errors
of 580.2\,km for IPinfo, 237.7\,km for MaxMind, 196.9\,km for
IP2Location, and 245.5\,km for DB-IP. These alignment differences add
information beyond decoded error: IPinfo also has the lowest decoded
median error (4.7\,km). Although IP2Location has lower median error
than MaxMind (6.1 versus 7.6\,km), it shows weaker agreement with its
raw GeoDB path.

Paths with more negative PCS and high error are where $\mathcal{R}$ is most
diagnostic. Among paths with mean error above 200\,km and PCS in the bottom
(most negative) PCS quartile, 44.9\% have $\mathcal{R}<0.2$; in the more
severe corner combining top-decile error with bottom (most negative) PCS
decile, the median
$\mathcal{R}$ falls to 0.188 and 51.9\% of pairs have
$\mathcal{R}<0.2$. The same pattern appears per vendor: among paths with
more negative PCS and high error, the median $\mathcal{R}$ is 0.343 for
IPinfo and 0.379 for MaxMind, compared with 0.076 for IP2Location and 0.192
for DB-IP.
Thus, when PCS is more negative and geolocation error is high, the
failure is often not that the HMM selected an arbitrary bad city; rather,
the observed RTT increments and the available geographic paths no longer
support a stable speed-of-light residual profile. In these regimes,
latency ceases to be a dependable proxy for geography, and PCS should be
treated as a warning about path interpretability rather than as a
calibrated confidence score.

\section{Evaluation}
\label{sec:evaluation}

We next evaluate whether the validation-set patterns persist at full
corpus scale. As described in \Cref{sec:datasets:traceroutes} and
\Cref{table:datasets}, the Evaluation Corpus is a held-out RIPE Atlas
window used for large-scale inference, PCS analysis, and path-alignment
diagnostics. The full-corpus PCS CDF uses the 410,643 traceroutes that
yield decoded paths for each GeoDB vendor. Unlike the validated subset
in \Cref{sec:validation}, this corpus does not provide active-ping
ground truth for every hop, so we do not interpret the following
distributions as direct accuracy measurements. Instead, we use PCS and
$\mathcal{R}$ to ask whether the same vendor-specific interpretability
patterns observed under validation also appear when \name{} is applied
to the full measurement workload.

\begin{figure*}[t]
    \centering
    \begin{subfigure}[t]{0.32\textwidth}
        \centering
        \includegraphics[width=\linewidth]{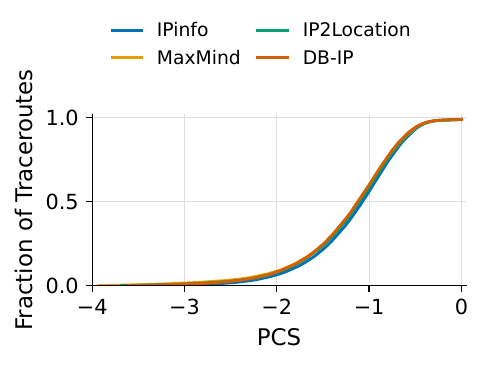}
        \caption{PCS distribution.}
        \label{fig:full-dataset-pcs-cdf}
    \end{subfigure}
    \hspace{0.08\textwidth}
    \begin{subfigure}[t]{0.32\textwidth}
        \centering
        \includegraphics[width=\linewidth]{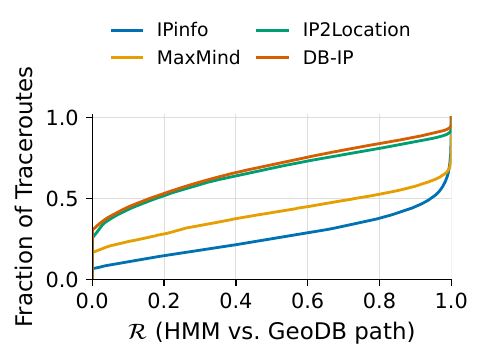}
        \caption{Alignment $\mathcal{R}$ distribution.}
        \label{fig:full-dataset-r-cdf}
    \end{subfigure}
    \caption{
        Decoded Evaluation Corpus: CDFs of PCS and Path--Model Alignment
        $\mathcal{R}$ across vendor-specific decoding runs. PCS is a
        log-score, so values closer to zero indicate stronger path
        consistency; $\mathcal{R}$ measures agreement between the HMM
        path and the raw GeoDB path, restricted to paths with at least
        two consecutive hops carrying both coordinates and RTT
        measurements in each path.
    }
    \label{fig:full-dataset-cdfs}
\end{figure*}

\Cref{fig:full-dataset-pcs-cdf} shows that full-corpus PCS distributions
are nearly identical across vendors. The median PCS ranges only from
-1.084 for IPinfo to -1.133 for DB-IP, and the 95th percentile ranges
from -0.480 to -0.508. Because PCS values closer to zero indicate
stronger path consistency, this overlap is an important negative result:
PCS is largely GeoDB-agnostic. Regardless of which commercial database
seeds the candidate set, the HMM converges to paths with similar model
fit once auxiliary evidence, endpoint anchors, and transition constraints
are considered. The lower tail is also shared across vendors. Between
7.4\% and 9.5\% of paths fall below PCS $<-2$, and the plotted CDFs omit
only the most extreme 1\% tail, whose cutoffs are already below
PCS $<-3.0$. Thus, PCS primarily measures the inherent reconstructability
of the observed traceroute under the HMM, not the quality of any
particular GeoDB trajectory.

\Cref{fig:full-dataset-r-cdf} exposes that missing dimension. In a CDF,
a lower curve is shifted toward larger $\mathcal{R}$ values; therefore
the IPinfo curve indicates the strongest path-model agreement. The
full-corpus ordering matches the validated-set conclusion in
\Cref{sec:validation:alignment_metric}: IPinfo is best, MaxMind is
second, and IP2Location and DB-IP have substantially larger
low-alignment mass. For IPinfo, the median $\mathcal{R}$ is 0.943,
62.1\% of measurable paths have $\mathcal{R}\ge0.8$, and 35.3\% are
near-perfect matches with $\mathcal{R}\ge0.99$. MaxMind remains
intermediate, with median $\mathcal{R}=0.729$ and 47.4\% of paths at
$\mathcal{R}\ge0.8$. By contrast, IP2Location and DB-IP have median
$\mathcal{R}$ values of 0.177 and 0.159, and only 19.0\% and 16.0\% of
paths reach $\mathcal{R}\ge0.8$.

The contrast between the two CDFs is the main full-corpus result. PCS
shows that \name{} often produces a coherent latency- and metadata-aware
path even when initialized from different GeoDB vendors; $\mathcal{R}$
shows whether that coherent HMM path is aligned with the raw vendor path.
This distinction mirrors the validation analysis: IPinfo and MaxMind
produce reference paths that more often agree with the HMM's
speed-of-light residual profile, whereas IP2Location and DB-IP more
often require the HMM to move away from the raw GeoDB trajectory to
obtain a plausible path. This effect is not confined to a small tail:
51.8\% of IP2Location paths and 53.1\% of DB-IP paths have
$\mathcal{R}<0.2$, compared with 14.8\% for IPinfo and 28.2\% for
MaxMind. Full-corpus evaluation therefore reinforces the validated-set
interpretation while extending it beyond the small actively validated
subset: vendor quality is not only a question of per-hop error, but also
of whether a database induces path-level geographies that remain
interpretable under traceroute latency constraints.

\subsection{Case Studies}
\label{sec:case-studies}

We ground the aggregate patterns above in four individual traceroutes
that span the alignment spectrum. Full per-vendor details are in
\Cref{sec:appendix:case-studies}.

\vspace{0.3em}
\noindent
\textbf{High-confidence alignment.}
A four-hop path from Auckland, NZ to Nuremberg, DE illustrates the
ideal case. IPinfo and IP2Location place every hop at the validated
location, yielding $\mathcal{R}=1.00$ and PCS\,$=-0.57$. DB-IP
assigns hop~1 to Lower Hutt instead of Auckland (34\,km away) but the
HMM corrects it, producing $\mathcal{R}=0.65$. The alignment gap
between vendors is visible even on a path this short.

\vspace{0.3em}
\noindent
\textbf{Bogon-aware decode.}
A Zurich--to--Indonesia path contains a bogon hop~7
(\texttt{10.111.222.210}), which no GeoDB can map. Two vendors
(DB-IP, IP2Location) place the adjacent responding hops~5--6 in
Amsterdam, yet the 145\,ms RTT from Zurich is consistent with Singapore,
not Amsterdam. The HMM moves hops~5--6 to Singapore and propagates
that context through the bogon hop, matching the validated locations.
The DB-IP run achieves $\mathcal{R}=1.00$; by contrast, IP2Location
yields $\mathcal{R}=0.05$ because its raw GeoDB path (Amsterdam for
hops~5--6) disagrees sharply with the decoded path.

\vspace{0.3em}
\noindent
\textbf{rDNS-supported correction.}
On a Netherlands--to--Minnesota path, DB-IP maps hop~8 to Singapore while
the rDNS record encodes Winnipeg. The HMM selects Winnipeg, matching
the active-ping validation. Despite this per-hop correction, the
overall path has PCS\,$=-1.51$ and low $\mathcal{R}$ across all vendors
($0.00$--$0.33$), because the rDNS correction is a local fix on a path
whose RTT profile does not smoothly track geographic distance.

\vspace{0.3em}
\noindent
\textbf{Potential MPLS tunnel.}
A Miami--to--Sydney path has only three visible hops. Hop~4 validates
to Miami, but its 231\,ms RTT from the source anchor is consistent
with Sydney, so the HMM moves it there. All vendors yield
$\mathcal{R}=0.00$ and PCS\,$=-1.63$. This is a deliberate failure
mode: the missing intermediate hops, likely hidden by an MPLS tunnel,
remove the latency structure that the transition model needs. The low
$\mathcal{R}$ and PCS correctly flag this path as one where the
geographic interpretation should not be trusted.

\section{Discussion}
\label{sec:discussion}

\name{} deliberately separates path inference from confidence scoring, but its
strongest correctness evidence comes from a small actively validated slice of
the Evaluation Corpus. The full Evaluation Corpus supports large-scale
decoding and PCS analysis, whereas only 6,555 traceroutes, or approximately
1.6\% of the corpus, have hop-level active-ping validation. This low validation
fraction is an explicit scope limit: the validation results measure behavior on
paths where ground-truth collection was feasible, not a direct accuracy
guarantee for every decoded path in the corpus.

\vspace{0.3em}
\noindent
\textbf{Validation Coverage.}
The 2\,ms shortest-ping rule provides a conservative local constraint, but it
does not validate every router interface in the Evaluation Corpus. Some
routers do not respond to active probes, some regions have sparse RIPE Atlas
coverage, and IXP or facility-adjacent addresses may be topologically
meaningful without behaving like ordinary router interfaces. \name{} handles
these cases by lowering evidence certainty and relying on path context, but
full-corpus findings should be interpreted as consistency evidence unless they
are restricted to the actively validated subset.

\vspace{0.3em}
\noindent
\textbf{Traceroute Latency.}
Traceroute RTT increments are not pure propagation measurements. Congestion,
asymmetric return paths, MPLS tunnels, router rate limiting, and ICMP
prioritization can all inflate or distort hop-level latency. For this reason,
\name{} uses soft transition penalties and empirical priors rather than a hard
speed-of-light cutoff. A more negative PCS should be interpreted as evidence
that the geographic interpretation is not well supported, not necessarily proof
that a specific database entry is wrong.

\vspace{0.3em}
\noindent
\textbf{Alignment Scope.}
The alignment metric $\mathcal{R}$ compares speed-of-light residual
increments between the decoded path and a reference path, but agreement is
not the same as correctness.
$\mathcal{R}$ cannot detect cases where both paths are consistently
wrong---for example, systematic database errors at locations that are
latency-consistent with neighboring hops.
Diagnosing \emph{why} alignment is low after the metric identifies a
path-model disagreement is a classification problem beyond the scope of a
single residual-profile metric and is left to future work.

\vspace{0.3em}
\noindent
\textbf{Use in Downstream Measurement.}
\name{} is most useful when downstream analyses need to know whether a
geographic interpretation is dependable. The decoded path can suggest likely
router locations, but the more robust output is the consistency score and the
segment-level explanation of where evidence and path constraints disagree.
At the dataset level, the distribution of $\mathcal{R}$ characterizes how
trustworthy the PCS scores are for a given reference path, and the
distribution of PCS itself characterizes how well the geolocation is
supported.
Reporting both distributions alongside geographic findings lets researchers
quantify the confidence in their conclusions rather than silently inheriting
database accuracy.
Future work should extend the same framework to unconstrained endpoints,
additional empirical-prior granularities, and longitudinal studies of how
router-location evidence changes over time. More broadly, more negative PCS
and low-$\mathcal{R}$ regions can identify where geolocation remains poorly
supported by current public evidence, indicating where additional active
measurements, RIPE Atlas probe deployment, or operator-published metadata would
most improve the trustworthiness of inferred router locations.

\section{Related Work}
\label{sec:related-work}

\vspace{0.3em}
\noindent
\textbf{Router Geolocation.}
Router geolocation research has produced several complementary ways to infer where an IP address or router interface is located.
Measurement-based methods use latency and topology observations to constrain feasible locations, from
IP-to-location mapping~\cite{geolocation:geo-mapping-techniques:ccr01}
and constraint-based geolocation~\cite{geolocation:cbg:imc04}
to topology-aware~\cite{geolocation:ethan-topology-geolocation:imc06}
and Internet-scale active systems~\cite{geolocation:towards-geo-million:imc12,geolocation:ben-ipmap:ccr20,geolocation:georesolver:conext25}.
These methods provide valuable location evidence, but they depend on contemporaneous probing, responsive targets, and well-placed vantage points;
recent replication work shows that accuracy and coverage claims can be difficult to reproduce with public measurement infrastructure~\cite{geolocation:replication-darwich:conext23}.
\name{} addresses a complementary problem: given an existing traceroute and the location evidence already available for its hops, it asks whether the resulting geographic path is internally consistent.

Passive router geolocation uses metadata that is easier to obtain at scale but less uniformly trustworthy.
Commercial geolocation databases provide broad coverage, yet prior studies show that their entries can be unreliable for Internet infrastructure and router interfaces~\cite{geolocation:geodb-unreliable:ccr11,geolocation:roya-geodb:imc17}.
Operator-published sources such as Geofeeds add useful hints, but they are not complete ground truth and can still contain erroneous router locations \cite{geolocation:ioana-geofeeds:conext24}.
Hostname and DNS PTR systems extract interpretable location hints from operator naming conventions, using
domain-specific rules~\cite{geolocation:drop:ccr14},
learned regular expressions~\cite{geolocation:hoiho:conext21}, or
language-models~\cite{geolocation:aleph:conext25}
to decode geographic labels.
These approaches improve geolocation for individual hops, whereas \name{} treats those as noisy observational-candidates in a path-level model that can reinforce consistent evidence and discount locally plausible but physically implausible assignments.

\vspace{0.3em}
\noindent
\textbf{Geolocation Quality and Confidence.}
Several lines of work characterize how accurate geolocation is, but none
provides a per-path confidence signal from passive data alone.
Measurement-based methods such as CBG~\cite{geolocation:cbg:imc04}
produce feasibility regions rather than point estimates, but they
require active probing from distributed vantage points and cannot be
applied retroactively to existing traceroute datasets.
Database comparison studies show that commercial GeoDBs disagree
substantially for infrastructure
addresses~\cite{geolocation:geodb-unreliable:ccr11,geolocation:roya-geodb:imc17},
yet their findings characterize population-level accuracy, not the
reliability of any individual path.
Geofeeds add operator-published evidence, but recent work shows they
are not uniformly correct either~\cite{geolocation:ioana-geofeeds:conext24}.
\name{} addresses the gap between per-hop accuracy studies and
path-level trust: it attaches a continuous consistency score to each
traceroute so that downstream analyses can quantify confidence in
their geographic conclusions rather than treating every mapping as
equally trustworthy.

\vspace{0.3em}
\noindent
\textbf{Path-Level Constraints in Traceroute.}
Traceroute provides sequential evidence that can constrain geographic
interpretations of router locations.
RTT increments and speed-of-light limits bound how far adjacent hops can
plausibly be from one another, while larger path patterns can expose
long-distance forwarding structure \cite{paxson96,singla2014internet}.
Prior work uses these signals to study delay-geography relationships and to
connect logical Internet paths with physical infrastructure, including
long-haul links~\cite{long-haul-links}, submarine-cable candidates
\cite{ramanathan:nautilus,calypso:sigcomm25}, and cross-layer maps of the
Internet~\cite{igdb-2022-imc}.
These studies show that traceroute contains useful geographic signal, but they
also target physical path discovery or infrastructure mapping.
\name{} uses the same class of path-level constraints for a different purpose:
to score whether a sequence of metadata-supported hop locations is internally
consistent.

\vspace{0.3em}
\noindent
\textbf{Traceroute Artifacts.}
Traceroute artifacts limit how directly path constraints can be interpreted as
geographic facts.
Delay anomalies and asymmetric forwarding can weaken the relationship between
observed RTT and physical distance~\cite{imc17:romain}, and MPLS deployments can
hide internal hops, expose only tunnel endpoints, or create IP-level adjacencies
that do not correspond cleanly to physical router adjacency
\cite{mpls-paul-2011-imc,matthew-mpls-2012-ccr,wormhole-2017-imc}.
These artifacts make low path consistency ambiguous: an implausible geographic
sequence may reflect incorrect geolocation metadata, but it may also reflect a
traceroute segment where the measurement no longer exposes the underlying
physical path.
\name{} therefore treats traceroute order and RTT increments as reliability
constraints rather than as a complete physical map, allowing PCS to identify
paths whose geographic interpretation should be trusted or treated with caution.

\vspace{0.3em}
\noindent
\textbf{Sequential Inference in Network Measurement.}
Hidden Markov Models and related Markov models are useful in network measurement when observable traffic or latency signals must be explained by an unobserved sequence of network states.
Prior work uses this structure to
segment Internet path-delay time series into latent delay regimes~\cite{hmm:characterization-delay-infinite-hmms:access20},
infer dominant congested links from end-to-end observations~\cite{hmm:model-based-identification:ton11},
and classify traffic~\cite{hmm:classification-using-hmm:globe08}
or protocol behavior~\cite{hmm:unexpected-means:imc06}
from packet- or flow-level sequences~\cite{hmm:tcp-classification-mm:tma10}.
These systems show that sequential models can recover structure that is not visible from any single observation, but their hidden states represent delay, congestion, or traffic classes rather than geographic hypotheses.
\name{} uses the same sequential-inference principle for path-level geolocation consistency where hidden states are city-level location candidates, emissions summarize support from router metadata, and transitions encode physical and empirical latency constraints.
The resulting model is not intended to introduce HMMs to network measurement; it uses an HMM as a scoring mechanism for when passive geolocation evidence, traceroute latency, and a decoded geographic sequence agree or disagree.

\section{Conclusion}
\label{sec:conclusion}

This paper addresses a common weakness in traceroute-based Internet
measurement: router locations are often used as point facts even when the
supporting evidence is sparse, conflicting, or inconsistent with the measured
path. \name{} reframes infrastructure geolocation as a path-level consistency
problem. By combining candidate locations from GeoDBs, rDNS, Geofeeds, and IXP
metadata with physical and empirical latency constraints, \name{} produces both
a decoded geographic path and a path consistency score.

The key idea is to ask whether the full sequence of router-location hypotheses
forms a coherent explanation of the traceroute, rather than asking which source
should win at each hop independently. A complementary alignment metric measures
whether the decoded path and a reference path produce consistent
speed-of-light residual increments, enabling dataset-level assessment of
when confidence scores are meaningful and when latency evidence is outside
the model scope.
On validated paths, 94.2\% of decoded sequences achieve mean error below
200\,km, and PCS is largely GeoDB-agnostic---median scores vary by less than
5\% across four commercial databases---while the alignment metric reveals
sharp vendor differences in how often the raw database path agrees with the
latency-constrained decode.
This formulation lets downstream
measurement studies separate well-supported geographic interpretations from
paths that require caution, and it exposes the segments where metadata and
latency evidence disagree. A current limitation is validation coverage:
fully verified paths depend on responsive routers and nearby RIPE Atlas probes.
Extending PCS to sparser regions, unconstrained endpoints, and longitudinal
measurements is an important next step.

\bibliographystyle{ACM-Reference-Format}
\bibliography{base,ref}

@string{sigcomm-ccr = "ACM SIGCOMM CCR"}

@string{sigcomm-ccr = "ACM SIGCOMM Computer Communication Review"}

@string{imc = "Proc. of IMC"}

@string{sigcomm = "Proc. of ACM SIGCOMM"}

@string{hotnets = "Proc. of HotNets"}

@string{tma = "Proc. of TMA"}

@string{conext = {Proc. of CoNEXT}}

@inproceedings{hayashibara2004phi,
  author    = {Hayashibara, Naohiro and Defago, Xavier and Yared, Rami and Katayama, Takuya},
  title     = {The $\varphi$ accrual failure detector},
  booktitle = {Proc. IEEE Symposium on Reliable Distributed Systems (SRDS)},
  year      = {2004},
  pages     = {66--78},
  publisher = {IEEE}
}

@article{corbett2013spanner,
  author    = {Corbett, James C. and Dean, Jeffrey and Epstein, Michael and Fikes, Andrew and Frost, Christopher and Furman, J. J. and Ghemawat, Sanjay and Gubarev, Andrey and Heiser, Christopher and Hochschild, Peter and others},
  title     = {Spanner: {Google}'s globally distributed database},
  journal   = {ACM Transactions on Computer Systems},
  volume    = {31},
  number    = {3},
  pages     = {1--22},
  year      = {2013},
  publisher = {ACM}
}

@inproceedings{aleph:conext,
  author    = {Kedar Thiagarajan and Esteban Carisimo  and  Fabián E. Bustamante},
  title     = {The Aleph: Decoding DNS PTR Records With Large Language Models},
  booktitle = {ACM CoNEXT},
  year      = {2025},
  month     = {12}
}

@inproceedings{paxson96,
  author    = {Vern Paxson},
  title     = {End-to-End Routing Behavior in the Internet},
  booktitle = sigcomm,
  year      = {1996}
}

@inproceedings{singla2014internet,
  title     = {The internet at the speed of light},
  author    = {Singla, Ankit and Chandrasekaran, Balakrishnan and Godfrey, P Brighten and Maggs, Bruce},
  booktitle = hotnets,
  year      = {2014}
}

@inproceedings{gharaibeh2017routergeo,
  title     = {A look at router geolocation in public and commercial databases},
  author    = {Gharaibeh, Manaf and Shah, Anant and Huffaker, Bradley and Zhang, Han and Ensafi, Roya and Papadopoulos, Christos},
  booktitle = imc,
  year      = {2017}
}

@inproceedings{imc-2023-geolocation-replication,
  author    = {Darwich, Omar and Rimlinger, Hugo and Dreyfus, Milo and Gouel, Matthieu and Vermeulen, Kevin},
  title     = {Replication: Towards a Publicly Available Internet Scale IP Geolocation Dataset},
  year      = {2023},
  isbn      = {9798400703829},
  publisher = {Association for Computing Machinery},
  address   = {New York, NY, USA},
  url       = {https://doi.org/10.1145/3618257.3624801},
  doi       = {10.1145/3618257.3624801},
  booktitle = {Proceedings of the 2023 ACM on Internet Measurement Conference},
  pages     = {1-15},
  numpages  = {15},
  location  = {Montreal QC, Canada},
  series    = {IMC '23}
}

@misc{IPinfo2025,
  title        = {IPinfo: IP Data Intelligence for Developers \& Enterprises},
  author       = {IPinfo, Inc.},
  howpublished = {https://ipinfo.io},
  year         = {2025},
  note         = {Accessed: 2025-11-05}
}

@article{pose:geodb,
  author       = {Ingmar Poese and Steve Uhlig and Mohamed Ali Kaafar and Benoit Donnet and Bamba Gueye},
  title        = {IP Geolocation Databases: Unreliable?},
  journaltitle = sigcomm-ccr,
  year         = {2011},
  volume       = {41},
  number       = {2},
  month        = {April}
}

@inproceedings{luckie:hoiho,
  author    = {Luckie, Matthew and Huffaker, Bradley and Marder, Alexander and Bischof, Zachary and Fletcher, Marianne and Claffy, K},
  title     = {Learning to extract geographic information from internet router hostnames},
  year      = {2021},
  booktitle = conext
}

@inproceedings{scheitle:hloc,
  author    = {Scheitle, Quirin and Gasser, Oliver and Sattler, Patrick and Carle, Georg},
  booktitle = tma,
  title     = {HLOC: Hints-based geolocation leveraging multiple measurement frameworks},
  year      = {2017}
}

@article{du2020ripe,
  title   = {RIPE IPmap active geolocation: Mechanism and performance evaluation},
  author  = {Du, Ben and Candela, Massimo and Huffaker, Bradley and Snoeren, Alex C and Claffy, KC},
  journal = sigcomm-ccr,
  year    = {2020}
}

@inproceedings{katz2006towards,
  author    = {Katz-Bassett, Ethan and John, John P. and Krishnamurthy, Arvind and Wetherall, David and Anderson, Thomas and Chawathe, Yatin},
  title     = {Towards IP geolocation using delay and topology measurements},
  year      = {2006},
  isbn      = {1595935614},
  publisher = {Association for Computing Machinery},
  address   = {New York, NY, USA},
  url       = {https://doi.org/10.1145/1177080.1177090},
  doi       = {10.1145/1177080.1177090},
  booktitle = {Proceedings of the 6th ACM SIGCOMM Conference on Internet Measurement},
  pages     = {71–84},
  numpages  = {14},
  location  = {Rio de Janeriro, Brazil},
  series    = {IMC '06}
}

@article{dong2012ip,
  title    = {Network measurement based modeling and optimization for IP geolocation},
  journal  = {Computer Networks},
  volume   = {56},
  number   = {1},
  pages    = {85-98},
  year     = {2012},
  issn     = {1389-1286},
  doi      = {https://doi.org/10.1016/j.comnet.2011.08.011},
  url      = {https://www.sciencedirect.com/science/article/pii/S1389128611003173},
  author   = {Ziqian Dong and Rohan D.W. Perera and Rajarathnam Chandramouli and K.P. Subbalakshmi}
}

@misc{peeringdb,
  author       = {{PeeringDB}},
  title        = {{PeeringDB}},
  howpublished = {\url{https://www.peeringdb.com/}},
  note         = {Dates used: 2023-12-08. Accessed: 2023-12-08},
  year         = 2023
}

@article{matthew-mpls-2012-ccr,
  author   = {Donnet, Benoit and Luckie, Matthew and M\'{e}rindol, Pascal and Pansiot, Jean-Jacques},
  title    = {Revealing MPLS tunnels obscured from traceroute},
  year     = {2012},
  volume   = {42},
  number   = {2},
  journal  = sigcomm-ccr,
  month    = mar,
  pages    = {87-93},
  numpages = {7}
}

@inproceedings{igdb-2022-imc,
  author    = {Anderson, Scott and Salamatian, Loqman and Bischof, Zachary S. and Dainotti, Alberto and Barford, Paul},
  title     = {iGDB: connecting the physical and logical layers of the internet},
  year      = {2022},
  isbn      = {9781450392594},
  publisher = {Association for Computing Machinery},
  address   = {New York, NY, USA},
  url       = {https://doi.org/10.1145/3517745.3561443},
  doi       = {10.1145/3517745.3561443},
  booktitle = {Proceedings of the 22nd ACM Internet Measurement Conference},
  pages     = {433–448},
  numpages  = {16},
  location  = {Nice, France},
  series    = {IMC '22}
}

@inproceedings{mpls-paul-2011-imc,
  author    = {Sommers, Joel and Barford, Paul and Eriksson, Brian},
  title     = {On the prevalence and characteristics of MPLS deployments in the open internet},
  year      = {2011},
  booktitle = imc
}

@inproceedings{wormhole-2017-imc,
  author    = {Vanaubel, Yves and M\'{e}rindol, Pascal and Pansiot, Jean-Jacques and Donnet, Benoit},
  title     = {Through the wormhole: tracking invisible MPLS tunnels},
  year      = {2017},
  booktitle = imc
}

@article{long-haul-links,
  author    = {Carisimo, Esteban and Wang, Caleb J. and Weaver, Mia and Bustamante, Fabi\'{a}n E. and Barford, Paul},
  title     = {A Hop Away from Everywhere: A View of the Intercontinental Long-haul Infrastructure},
  year      = {2023},
  journal   = {Proc. ACM Meas. Anal. Comput. Syst.},
  month     = {dec},
  articleno = {47},
  numpages  = {26}
}

@article{rabiner1986,
  author  = {Rabiner, Lawrence R. and Juang, Biing-Hwang},
  title   = {An Introduction to Hidden Markov Models},
  journal = {IEEE ASSP Magazine},
  volume  = {3},
  number  = {1},
  pages   = {4--16},
  year    = {1986},
  doi     = {10.1109/MASSP.1986.1165342}
}

@techreport{bilmes1998,
  author      = {Bilmes, Jeff A.},
  title       = {A Gentle Tutorial of the EM Algorithm and Its Application to Parameter Estimation for Gaussian Mixture and Hidden Markov Models},
  institution = {International Computer Science Institute},
  year        = {1998},
  url         = {https://www.cs.cmu.edu/~aarti/Class/10701/readings/gentle_tut_HMM.pdf}
}

@article{ephraim2002,
  author  = {Ephraim, Yariv and Merhav, Neri},
  title   = {Hidden Markov Processes},
  journal = {IEEE Transactions on Information Theory},
  volume  = {48},
  number  = {6},
  pages   = {1518--1569},
  year    = {2002},
  doi     = {10.1109/TIT.2002.1003838}
}

@misc{IPinfoBogon2025,
  title        = {Bogon IP Address Ranges},
  author       = {IPinfo, Inc.},
  howpublished = {https://ipinfo.io/bogon},
  year         = {2025},
  note         = {Accessed: 2025-11-05}
}

@misc{CAIDA_Bogon2025,
  title        = {CYMRU Bogon Reference Dataset (historical and daily bogons and fullbogons) - Center for Applied Internet Data Analysis (CAIDA) / Team Cymru},
  author       = {Center for Applied Internet Data Analysis (CAIDA)},
  howpublished = {https://publicdata.caida.org/datasets/bogon/},
  year         = {2025},
  note         = {Accessed: 2025-11-05}
}

@inproceedings{imc17:romain,
  title     = {Pinpointing delay and forwarding anomalies using
               large-scale traceroute measurements},
  author    = {Romain Fontugne and Cristel Pelsser and Emile Aben
               and Randy Bush},
  booktitle = imc,
  year      = 2017,
  month     = {November}
}

@misc{geolocatemuch,
  author       = {Massimo Candela, Emanuele Candela, Lorenzo Ariemma},
  title        = {Geofeeds Registry "Geolocate much?"},
  howpublished = {\url{https://geolocatemuch.com/}},
  year         = {2025}
}

@misc{maxmindgeoloc,
  author       = {Maxmind},
  title        = {Maxmind Geolocation Data},
  howpublished = {\url{https://www.maxmind.com/en/geoip2-services-and-databases}},
  year         = {2025}
}

@misc{dbipgeoloc,
  author       = {DB-IP},
  title        = {DB-IP Geolocation Data},
  howpublished = {\url{https://db-ip.com/db/}},
  year         = {2025}
}

@misc{ip2locationgeoloc,
  author       = {IP2Location},
  title        = {IP2Location Geolocation Data},
  howpublished = {\url{https://www.ip2location.com/databases}},
  year         = {2025}
}

@article{ramanathan:nautilus,
  author  = {Alagappan Ramanathan and Sangeetha Abdu Jyothi},
  title   = {Nautilus: A Framework for Cross-Layer Cartography of Submarine Cables and IP Links},
  year    = {2023},
  volume  = {7},
  number  = {3},
  journal = {Proc. ACM Meas. Anal. Comput. Syst.},
  month   = {dec}
}

@misc{ripe2017anchoring,
  author       = {Davies, Alun},
  title        = {Anchoring Measurements: Bringing Back the Balance},
  howpublished = {\url{https://labs.ripe.net/author/alun_davies/anchoring-measurements-bringing-back-the-balance/}},
  year         = {2017},
  month        = {June},
  note         = {RIPE Labs article}
}

@article{geolocation:geo-mapping-techniques:ccr01,
  author     = {Padmanabhan, Venkata N. and Subramanian, Lakshminarayanan},
  title      = {An investigation  of geographic mapping techniques for internet hosts},
  year       = {2001},
  issue_date = {October 2001},
  publisher  = {Association for Computing Machinery},
  address    = {New York, NY, USA},
  volume     = {31},
  number     = {4},
  issn       = {0146-4833},
  url        = {https://doi.org/10.1145/964723.383073},
  doi        = {10.1145/964723.383073},
  journal    = {SIGCOMM Comput. Commun. Rev.},
  month      = aug,
  pages      = {173–185},
  numpages   = {13}
}

@inproceedings{geolocation:cbg:imc04,
  author    = {Gueye, Bamba and Ziviani, Artur and Crovella, Mark and Fdida, Serge},
  title     = {Constraint-based geolocation of internet hosts},
  year      = {2004},
  isbn      = {1581138210},
  publisher = {Association for Computing Machinery},
  address   = {New York, NY, USA},
  url       = {https://doi.org/10.1145/1028788.1028828},
  doi       = {10.1145/1028788.1028828},
  booktitle = {Proceedings of the 4th ACM SIGCOMM Conference on Internet Measurement},
  pages     = {288–293},
  numpages  = {6},
  location  = {Taormina, Sicily, Italy},
  series    = {IMC '04}
}

@inproceedings{geolocation:ethan-topology-geolocation:imc06,
  author    = {Katz-Bassett, Ethan and John, John P. and Krishnamurthy, Arvind and Wetherall, David and Anderson, Thomas and Chawathe, Yatin},
  title     = {Towards IP geolocation using delay and topology measurements},
  year      = {2006},
  isbn      = {1595935614},
  publisher = {Association for Computing Machinery},
  address   = {New York, NY, USA},
  url       = {https://doi.org/10.1145/1177080.1177090},
  doi       = {10.1145/1177080.1177090},
  booktitle = {Proceedings of the 6th ACM SIGCOMM Conference on Internet Measurement},
  pages     = {71–84},
  numpages  = {14},
  location  = {Rio de Janeriro, Brazil},
  series    = {IMC '06}
}

@inproceedings{geolocation:towards-geo-million:imc12,
  author    = {Hu, Zi and Heidemann, John and Pradkin, Yuri},
  title     = {Towards geolocation of millions of IP addresses},
  year      = {2012},
  isbn      = {9781450317054},
  publisher = {Association for Computing Machinery},
  address   = {New York, NY, USA},
  url       = {https://doi.org/10.1145/2398776.2398790},
  doi       = {10.1145/2398776.2398790},
  booktitle = {Proceedings of the 2012 Internet Measurement Conference},
  pages     = {123–130},
  numpages  = {8},
  location  = {Boston, Massachusetts, USA},
  series    = {IMC '12}
}

@article{geolocation:geodb-unreliable:ccr11,
  author     = {Poese, Ingmar and Uhlig, Steve and Kaafar, Mohamed Ali and Donnet, Benoit and Gueye, Bamba},
  title      = {IP geolocation databases: unreliable?},
  year       = {2011},
  issue_date = {April 2011},
  publisher  = {Association for Computing Machinery},
  address    = {New York, NY, USA},
  volume     = {41},
  number     = {2},
  issn       = {0146-4833},
  url        = {https://doi.org/10.1145/1971162.1971171},
  doi        = {10.1145/1971162.1971171},
  journal    = {SIGCOMM Comput. Commun. Rev.},
  month      = apr,
  pages      = {53–56},
  numpages   = {4}
}

@article{geolocation:drop:ccr14,
  author     = {Huffaker, Bradley and Fomenkov, Marina and claffy, kc},
  title      = {DRoP: DNS-based router positioning},
  year       = {2014},
  issue_date = {July 2014},
  publisher  = {Association for Computing Machinery},
  address    = {New York, NY, USA},
  volume     = {44},
  number     = {3},
  issn       = {0146-4833},
  url        = {https://doi.org/10.1145/2656877.2656879},
  doi        = {10.1145/2656877.2656879},
  journal    = {SIGCOMM Comput. Commun. Rev.},
  month      = jul,
  pages      = {5–13},
  numpages   = {9}
}

@inproceedings{geolocation:roya-geodb:imc17,
  author    = {Gharaibeh, Manaf and Shah, Anant and Huffaker, Bradley and Zhang, Han and Ensafi, Roya and Papadopoulos, Christos},
  title     = {A look at router geolocation in public and commercial databases},
  year      = {2017},
  isbn      = {9781450351188},
  publisher = {Association for Computing Machinery},
  address   = {New York, NY, USA},
  url       = {https://doi.org/10.1145/3131365.3131380},
  doi       = {10.1145/3131365.3131380},
  booktitle = {Proceedings of the 2017 Internet Measurement Conference},
  pages     = {463–469},
  numpages  = {7},
  location  = {London, United Kingdom},
  series    = {IMC '17}
}

@inproceedings{geolocation:hoiho:conext21,
  author    = {Luckie, Matthew and Huffaker, Bradley and Marder, Alexander and Bischof, Zachary and Fletcher, Marianne and Claffy, K},
  title     = {Learning to extract geographic information from internet router hostnames},
  year      = {2021},
  isbn      = {9781450390989},
  publisher = {Association for Computing Machinery},
  address   = {New York, NY, USA},
  url       = {https://doi.org/10.1145/3485983.3494869},
  doi       = {10.1145/3485983.3494869},
  booktitle = {Proceedings of the 17th International Conference on Emerging Networking EXperiments and Technologies},
  pages     = {440–453},
  numpages  = {14},
  location  = {Virtual Event, Germany},
  series    = {CoNEXT '21}
}

@inproceedings{geolocation:replication-darwich:conext23,
  author    = {Darwich, Omar and Rimlinger, Hugo and Dreyfus, Milo and Gouel, Matthieu and Vermeulen, Kevin},
  title     = {Replication: Towards a Publicly Available Internet Scale IP Geolocation Dataset},
  year      = {2023},
  isbn      = {9798400703829},
  publisher = {Association for Computing Machinery},
  address   = {New York, NY, USA},
  url       = {https://doi.org/10.1145/3618257.3624801},
  doi       = {10.1145/3618257.3624801},
  booktitle = {Proceedings of the 2023 ACM on Internet Measurement Conference},
  pages     = {1–15},
  numpages  = {15},
  location  = {Montreal QC, Canada},
  series    = {IMC '23}
}

@article{geolocation:ioana-geofeeds:conext24,
  author     = {Livadariu, Ioana and Vermeulen, Kevin and Mouchet, Maxime and Giotsas, Vasilis},
  title      = {Geofeeds: Revolutionizing IP Geolocation or Illusionary Promises?},
  year       = {2024},
  issue_date = {September 2024},
  publisher  = {Association for Computing Machinery},
  address    = {New York, NY, USA},
  volume     = {2},
  number     = {CoNEXT3},
  url        = {https://doi.org/10.1145/3676869},
  doi        = {10.1145/3676869},
  journal    = {Proc. ACM Netw.},
  month      = aug,
  articleno  = {15},
  numpages   = {21}
}

@article{geolocation:aleph:conext25,
  author     = {Thiagarajan, Kedar and Carisimo, Esteban and Bustamante, Fabi\'{a}n E.},
  title      = {The Aleph: Decoding Geographic Information from DNS PTR Records Using Large Language Models},
  year       = {2025},
  issue_date = {March 2025},
  publisher  = {Association for Computing Machinery},
  address    = {New York, NY, USA},
  volume     = {3},
  number     = {CoNEXT1},
  url        = {https://doi.org/10.1145/3709374},
  doi        = {10.1145/3709374},
  journal    = {Proc. ACM Netw.},
  month      = mar,
  articleno  = {7},
  numpages   = {20}
}

@article{geolocation:georesolver:conext25,
  author     = {Rimlinger, Hugo and Fourmaux, Olivier and Friedman, Timur and Vermeulen, Kevin},
  title      = {GeoResolver: An Accurate, Scalable, and Explainable Geolocation Technique Using DNS Redirection},
  year       = {2025},
  issue_date = {September 2025},
  publisher  = {Association for Computing Machinery},
  address    = {New York, NY, USA},
  volume     = {3},
  number     = {CoNEXT3},
  url        = {https://doi.org/10.1145/3749219},
  doi        = {10.1145/3749219},
  journal    = {Proc. ACM Netw.},
  month      = sep,
  articleno  = {19},
  numpages   = {21}
}

@article{geolocation:ben-ipmap:ccr20,
  author     = {Du, Ben and Candela, Massimo and Huffaker, Bradley and Snoeren, Alex C. and claffy, kc},
  title      = {RIPE IPmap active geolocation: mechanism and performance evaluation},
  year       = {2020},
  issue_date = {April 2020},
  publisher  = {Association for Computing Machinery},
  address    = {New York, NY, USA},
  volume     = {50},
  number     = {2},
  issn       = {0146-4833},
  url        = {https://doi.org/10.1145/3402413.3402415},
  doi        = {10.1145/3402413.3402415},
  journal    = {SIGCOMM Comput. Commun. Rev.},
  month      = may,
  pages      = {3–10},
  numpages   = {8}
}

@article{hmm:characterization-delay-infinite-hmms:access20,
  author   = {Mouchet, Maxime and Vaton, Sandrine and Chonavel, Thierry and Aben, Emile and Hertog, Jasper Den},
  journal  = {IEEE Access},
  title    = {Large-Scale Characterization and Segmentation of Internet Path Delays With Infinite HMMs},
  year     = {2020},
  volume   = {8},
  number   = {},
  pages    = {16771-16784},
  doi      = {10.1109/ACCESS.2020.2968380}
}

@article{hmm:model-based-identification:ton11,
  author   = {Wei, Wei and Wang, Bing and Towsley, Don and Kurose, Jim},
  journal  = {IEEE/ACM Transactions on Networking},
  title    = {Model-Based Identification of Dominant Congested Links},
  year     = {2011},
  volume   = {19},
  number   = {2},
  pages    = {456-469},
  doi      = {10.1109/TNET.2010.2068058}
}

@inproceedings{hmm:unexpected-means:imc06,
  author    = {Ma, Justin and Levchenko, Kirill and Kreibich, Christian and Savage, Stefan and Voelker, Geoffrey M.},
  title     = {Unexpected means of protocol inference},
  year      = {2006},
  isbn      = {1595935614},
  publisher = {Association for Computing Machinery},
  address   = {New York, NY, USA},
  url       = {https://doi.org/10.1145/1177080.1177123},
  doi       = {10.1145/1177080.1177123},
  booktitle = {Proceedings of the 6th ACM SIGCOMM Conference on Internet Measurement},
  pages     = {313–326},
  numpages  = {14},
  location  = {Rio de Janeriro, Brazil},
  series    = {IMC '06}
}

@inproceedings{hmm:classification-using-hmm:globe08,
  author    = {Dainotti, Alberto and de Donato, Walter and Pescape, Antonio and Salvo Rossi, Pierluigi},
  booktitle = {IEEE GLOBECOM 2008 - 2008 IEEE Global Telecommunications Conference},
  title     = {Classification of Network Traffic via Packet-Level Hidden Markov Models},
  year      = {2008},
  volume    = {},
  number    = {},
  pages     = {1-5},
  doi       = {10.1109/GLOCOM.2008.ECP.412}
}

@inproceedings{calypso:sigcomm25,
  author    = {Wang, Caleb and Zhang, Ying and Dong, Qianli and Carisimo, Esteban and Durairajan, Ramakrishnan and Bustamante, Fabi\'{a}n E.},
  title     = {Threading the Ocean: Mapping Digital Routes Across Submarine Cables using Calypso},
  year      = {2025},
  isbn      = {9798400715242},
  publisher = {Association for Computing Machinery},
  address   = {New York, NY, USA},
  url       = {https://doi.org/10.1145/3718958.3750512},
  doi       = {10.1145/3718958.3750512},
  booktitle = {Proceedings of the ACM SIGCOMM 2025 Conference},
  pages     = {1260–1262},
  numpages  = {3},
  location  = {S\~{a}o Francisco Convent, Coimbra, Portugal},
  series    = {SIGCOMM '25}
}

@inproceedings{hmm:tcp-classification-mm:tma10,
  author    = {M{\"u}nz, Gerhard
               and Dai, Hui
               and Braun, Lothar
               and Carle, Georg},
  editor    = {Ricciato, Fabio
               and Mellia, Marco
               and Biersack, Ernst},
  title     = {TCP Traffic Classification Using Markov Models},
  booktitle = {Traffic Monitoring and Analysis},
  year      = {2010},
  publisher = {Springer Berlin Heidelberg},
  address   = {Berlin, Heidelberg},
  pages     = {127--140},
  isbn      = {978-3-642-12365-8}
}

\appendix
\section{Ethics}
\label{sec:appendix-ethics}

\setlength{\floatsep}{0.6\baselineskip}
\setlength{\textfloatsep}{0.6\baselineskip}
\setlength{\intextsep}{0.6\baselineskip}

This work does not raise any ethical issues.

\section{Model Parameters}
\label{sec:appendix:parameters}

We use one fixed parameter configuration for all validation and evaluation
decodes. \Cref{tab:appendix:hmm-parameters,tab:appendix:transition-parameters}
list the emission, transition, revisit, and Path--Model Alignment parameters
introduced in \Cref{sec:methodology}. Tunable values were selected on the
validated traceroutes with a deterministic 60/20/20 train/test/holdout split;
the resulting configuration is held fixed across GeoDB vendors.

\begin{table*}[!t]
    \centering
    \caption{HMM path-alignment configuration used in all decoding runs.}
    \label{tab:appendix:hmm-parameters}
    \scriptsize
    \setlength{\tabcolsep}{4pt}
    \renewcommand{\arraystretch}{0.88}
    \begin{tabular}{p{0.14\textwidth}p{0.38\textwidth}p{0.28\textwidth}p{0.12\textwidth}}
        \toprule
        \textbf{Parameter} & \textbf{Description} & \textbf{Introduced in} & \textbf{Value} \\
        \midrule
        $w_{\mathrm{src}}, w_{\mathrm{dst}}$
            & Emission weight for endpoint anchors.
            & Endpoint anchors and emission utility in \Cref{sec:methodology:states,sec:methodology:emissions}.
            & 2 \\
        $w_{\mathrm{GeoDB}}$
            & Emission weight assigned to the selected GeoDB coordinate.
              Set to zero so that the GeoDB location enters only as a
              candidate state; the model selects or discards it based on
              transition constraints and agreement with auxiliary sources,
              ensuring PCS is not biased toward the GeoDB's own answer.
            & Candidate evidence and emission utility in \Cref{sec:methodology:states,sec:methodology:emissions}.
            & 0 \\
        $w_{\mathrm{Aleph\ rDNS}}$
            & Emission weight for Aleph rDNS hints.
            & Evidence-source weight $w_k$ in \Cref{sec:methodology:emissions}.
            & 0.4 \\
        $w_{\mathrm{Geofeed}}$
            & Emission weight for Geofeed candidates.
            & Evidence-source weight $w_k$ in \Cref{sec:methodology:emissions}.
            & 0.3 \\
        $w_{\mathrm{IXP}}$
            & Emission weight for IXP candidates.
            & Evidence-source weight $w_k$ in \Cref{sec:methodology:emissions}.
            & 0.6 \\
        $w_{\mathrm{peering}}$
            & Emission weight for inter-AS peering-location candidates.
            & Evidence-source weight $w_k$ in \Cref{sec:methodology:emissions}.
            & 0.3 \\
        $\tau_t$
            & Temperature in the emission softmax.
            & Emission distribution in \Cref{sec:methodology:emissions}.
            & $0.2U_{\max}$ \\
        $\gamma$
            & Uniform emission smoothing floor.
            & Smoothed emission distribution in \Cref{sec:methodology:emissions}.
            & 0.05 \\
        $\alpha_t$
            & Hop-specific certainty weight on emission evidence.
            & Evidence-certainty weighting in \Cref{sec:methodology:emissions}.
            & $[0,1]$ \\
        $\beta$
            & Transition path stiffness in the Viterbi objective.
            & Viterbi recurrence in \Cref{sec:methodology:scoring}.
            & 1 \\
        \bottomrule
    \end{tabular}
\end{table*}

\begin{table*}[!t]
    \centering
    \caption{Path-alignment transition, revisit, and Path--Model Alignment parameters.}
    \label{tab:appendix:transition-parameters}
    \scriptsize
    \setlength{\tabcolsep}{4pt}
    \renewcommand{\arraystretch}{0.86}
    \begin{tabular}{p{0.14\textwidth}p{0.38\textwidth}p{0.28\textwidth}p{0.12\textwidth}}
        \toprule
        \textbf{Parameter} & \textbf{Description} & \textbf{Introduced in} & \textbf{Value} \\
        \midrule
        $c$
            & Speed of light used by the transition model.
            & Propagation bound $\Delta_{\min}$ in \Cref{sec:methodology:transitions}.
            & 299.79\,km/ms \\
        $\rho$
            & Effective fiber-speed fraction.
            & Propagation bound $\Delta_{\min}$ in \Cref{sec:methodology:transitions}.
            & 0.66 \\
        $v=\rho c$
            & Effective propagation speed used for speed-of-light calculations.
            & Propagation bound $\Delta_{\min}$ in \Cref{sec:methodology:transitions}.
            & 197.86\,km/ms \\
        $\epsilon$
            & Residual floor in the physical slack term.
            & Residual slack $r_t$ in \Cref{sec:methodology:transitions}.
            & $10^{-12}$ \\
        $m$
            & Pareto-II residual scale.
            & Physical slack model in \Cref{sec:methodology:transitions}.
            & 20\,ms \\
        $\eta$
            & Pareto-II residual-shape parameter.
            & Physical slack model in \Cref{sec:methodology:transitions}.
            & 2.5 \\
        $k$
            & Logistic feasibility-gate slope.
            & Feasibility gate $g_t$ in \Cref{sec:methodology:transitions}.
            & 8 \\
        $K_{\lambda}$
            & Empirical-trust scale in $\lambda=N/(N+K_{\lambda})$.
            & Trust-weighted blend $\lambda_t$ in \Cref{sec:methodology:transitions}.
            & 50 samples \\
        $\lambda_{\max}$
            & Maximum empirical-transition weight.
            & Trust-weighted blend $\lambda_t$ in \Cref{sec:methodology:transitions}.
            & 0.85 \\
        $\lambda_{\mathrm{intra}}$
            & Empirical-transition override for same-country transitions.
            & Trust-weighted blend $\lambda_t$ in \Cref{sec:methodology:transitions}.
            & 0.2 \\
        $\sigma$
            & Gaussian smoothing bandwidth for empirical country-pair latency priors.
            & Empirical latency evidence in \Cref{sec:methodology:transitions}.
            & 2 \\
        $b_{\mathrm{KDE}}$
            & RTT-increment bin width used before kernel smoothing.
            & Empirical latency evidence in \Cref{sec:methodology:transitions}.
            & 5\,ms \\
        $b_{\mathrm{stay}}$
            & Base co-location bonus for consecutive hops decoded to the same city.
            & Co-location bonus in \Cref{sec:methodology:transitions}.
            & 0.4 \\
        $d_{\mathrm{stay}}$
            & RTT-decay scale for the co-location bonus.
            & Co-location bonus in \Cref{sec:methodology:transitions}.
            & 7\,ms \\
        $b_{\mathrm{rev}}$
            & Base additive log-penalty for revisits.
            & Revisit penalty in \Cref{sec:methodology:transitions}.
            & 1.6 \\
        $d_{\mathrm{rev}}$
            & Hop-distance decay for the revisit penalty.
            & Revisit penalty in \Cref{sec:methodology:transitions}.
            & 2 hops \\
        $g_{\min}$
            & Minimum intervening-hop gap required to classify a revisit.
            & Revisit penalty in \Cref{sec:methodology:transitions}.
            & 1 hop \\
        $v_{\mathcal{R}}$
            & Propagation speed used for Path--Model Alignment residuals.
            & Alignment propagation time $\hat{\Delta}_t$ in \Cref{sec:alignment}.
            & 200\,km/ms \\
        $\tau$
            & Soft floor in the normalized residual denominator.
            & Alignment residual $e_t(x)$ in \Cref{sec:alignment}.
            & 5\,ms \\
        $\epsilon_{\mathcal{R}}$
            & Numerical denominator floor in \Cref{eq:alignment}.
            & Path--Model Alignment score in \Cref{eq:alignment}.
            & $10^{-9}$ \\
        \bottomrule
    \end{tabular}
\end{table*}

\section{Case Study Details}
\label{sec:appendix:case-studies}

\Cref{tab:appendix:case-studies,tab:appendix:case-studies-continued} provide
full per-vendor hop tables for the four case studies discussed in
\Cref{sec:case-studies}.

\begin{table*}[!t]
    \centering
    \caption{Traceroute case studies. Hop RTT is the observed
    traceroute RTT in ms, Bogon is marked with \ding{51} only for bogon IPs,
    rDNS and IXP report public-source location evidence when available, and
    Validated loc. is the active validation location when available. Other
    public sources are omitted in these cases because they do not provide
    location mappings for the shown hops. PCS is a normalized log-score, so
    values closer to zero indicate a better HMM fit; $\mathcal{R}$ measures
    agreement between the HMM path and the GeoDB reference path.}
    \label{tab:appendix:case-studies}
    \tiny
    \setlength{\tabcolsep}{2.2pt}
    \renewcommand{\arraystretch}{0.86}
    \resizebox{\textwidth}{!}{%
    \begin{tabular}{@{}lclcrrlllllrr@{}}
        \toprule
        \textbf{Vendor} & \textbf{Hop} & \textbf{Hop IP} & \textbf{Bgn.} & \textbf{ASN} & \textbf{Hop RTT} & \textbf{rDNS} & \textbf{IXP} & \textbf{Validated loc.} & \textbf{GeoDB loc.} & \textbf{HMM loc.} & \textbf{PCS} & \textbf{$\mathcal{R}$} \\
        \midrule
        \multicolumn{13}{@{}l@{}}{\textbf{High-confidence alignment.} Source anchor 3112; destination anchor 2799; measurement 29988329; timestamp 2025-08-01 00:05:19; IPinfo/IP2Location yield $\mathcal{R}\approx1$.} \\
        \midrule
        \multirow{4}{*}{DB-IP} & 1 & \texttt{103.242.68.65} & & 133075 & 0.00 & -- & -- & Auckland, NZ & Lower Hutt, NZ & Auckland, NZ & \multirow{4}{*}{-0.60} & \multirow{4}{*}{0.65} \\
        & 2 & \texttt{131.203.72.165} & & 9790 & 1.00 & -- & -- & Auckland, NZ & Auckland (Auckland CBD), NZ & Auckland, NZ & & \\
        & 3 & \texttt{131.203.72.166} & & 9790 & 2.00 & -- & -- & Auckland, NZ & Auckland (Auckland CBD), NZ & Auckland, NZ & & \\
        & 5 & \texttt{193.27.55.25} & & 15451 & 286.00 & -- & -- & Nuremberg, DE & Nuremberg (Mitte), DE & Nuremberg, DE & & \\
        \cmidrule(lr){1-13}
        \multirow{4}{*}{IP2Location} & 1 & \texttt{103.242.68.65} & & 133075 & 0.00 & -- & -- & Auckland, NZ & Auckland, NZ & Auckland, NZ & \multirow{4}{*}{-0.57} & \multirow{4}{*}{1.00} \\
        & 2 & \texttt{131.203.72.165} & & 9790 & 1.00 & -- & -- & Auckland, NZ & Auckland, NZ & Auckland, NZ & & \\
        & 3 & \texttt{131.203.72.166} & & 9790 & 2.00 & -- & -- & Auckland, NZ & Auckland, NZ & Auckland, NZ & & \\
        & 5 & \texttt{193.27.55.25} & & 15451 & 286.00 & -- & -- & Nuremberg, DE & Nuremberg, DE & Nuremberg, DE & & \\
        \cmidrule(lr){1-13}
        \multirow{4}{*}{IPinfo} & 1 & \texttt{103.242.68.65} & & 133075 & 0.00 & -- & -- & Auckland, NZ & Auckland, NZ & Auckland, NZ & \multirow{4}{*}{-0.57} & \multirow{4}{*}{1.00} \\
        & 2 & \texttt{131.203.72.165} & & 9790 & 1.00 & -- & -- & Auckland, NZ & Auckland, NZ & Auckland, NZ & & \\
        & 3 & \texttt{131.203.72.166} & & 9790 & 2.00 & -- & -- & Auckland, NZ & Auckland, NZ & Auckland, NZ & & \\
        & 5 & \texttt{193.27.55.25} & & 15451 & 286.00 & -- & -- & Nuremberg, DE & Nuremberg, DE & Nuremberg, DE & & \\
        \cmidrule(lr){1-13}
        \multirow{4}{*}{MaxMind} & 1 & \texttt{103.242.68.65} & & 133075 & 0.00 & -- & -- & Auckland, NZ & Auckland, NZ & Auckland, NZ & \multirow{4}{*}{-0.60} & \multirow{4}{*}{0.33} \\
        & 2 & \texttt{131.203.72.165} & & 9790 & 1.00 & -- & -- & Auckland, NZ & Papakura, NZ & Auckland, NZ & & \\
        & 3 & \texttt{131.203.72.166} & & 9790 & 2.00 & -- & -- & Auckland, NZ & Papakura, NZ & Auckland, NZ & & \\
        & 5 & \texttt{193.27.55.25} & & 15451 & 286.00 & -- & -- & Nuremberg, DE & Nuremberg, DE & Nuremberg, DE & & \\
        \midrule
        \multicolumn{13}{@{}l@{}}{\textbf{Bogon-aware decoding.} Source anchor 1410; destination anchor 3807; measurement 70496250; timestamp 2025-08-01 00:13:48; hop 7 is bogon and decoded through Singapore.} \\
        \midrule
        \multirow{8}{*}{DB-IP} & 1 & \texttt{194.242.34.201} & & 20612 & 0.00 & -- & -- & Zurich, CH & Opfikon, CH & Glattbrugg, CH & \multirow{8}{*}{-0.82} & \multirow{8}{*}{1.00} \\
        & 2 & \texttt{91.206.52.37} & & 0 & 1.00 & -- & SwissIX, CH & -- & Zurich (Kreis 9), CH & Zurich, CH & & \\
        & 3 & \texttt{82.220.13.18} & & 9044 & 1.00 & Zurich, CH & -- & Zurich, CH & Solothurn, CH & Zurich, CH & & \\
        & 4 & \texttt{91.206.52.143} & & 0 & 0.00 & -- & SwissIX, CH & -- & Zurich (Kreis 9), CH & Zurich, CH & & \\
        & 5 & \texttt{87.245.234.83} & & 9002 & 145.00 & -- & -- & Singapore, SG & Amsterdam, NL & Singapore, SG & & \\
        & 6 & \texttt{87.245.231.201} & & 9002 & 144.00 & -- & -- & Singapore, SG & Amsterdam, NL & Singapore, SG & & \\
        & 7 & \texttt{10.111.222.210} & \ding{51} & 0 & 145.00 & -- & -- & -- & -- & Singapore, SG & & \\
        & 10 & \texttt{206.237.97.100} & & 140443 & 158.00 & -- & -- & Serang, ID & Cikarang, ID & Cibitung, ID & & \\
        \cmidrule(lr){1-13}
        \multirow{8}{*}{IP2Location} & 1 & \texttt{194.242.34.201} & & 20612 & 0.00 & -- & -- & Zurich, CH & Zurich, CH & Glattbrugg, CH & \multirow{8}{*}{-0.81} & \multirow{8}{*}{0.05} \\
        & 2 & \texttt{91.206.52.37} & & 0 & 1.00 & -- & SwissIX, CH & -- & Zurich, CH & Zurich, CH & & \\
        & 3 & \texttt{82.220.13.18} & & 9044 & 1.00 & Zurich, CH & -- & Zurich, CH & Bern, CH & Zurich, CH & & \\
        & 4 & \texttt{91.206.52.143} & & 0 & 0.00 & -- & SwissIX, CH & -- & Zurich, CH & Zurich, CH & & \\
        & 5 & \texttt{87.245.234.83} & & 9002 & 145.00 & -- & -- & Singapore, SG & Amsterdam, NL & Singapore, SG & & \\
        & 6 & \texttt{87.245.231.201} & & 9002 & 144.00 & -- & -- & Singapore, SG & Amsterdam, NL & Singapore, SG & & \\
        & 7 & \texttt{10.111.222.210} & \ding{51} & 0 & 145.00 & -- & -- & -- & -- & Singapore, SG & & \\
        & 10 & \texttt{206.237.97.100} & & 140443 & 158.00 & -- & -- & Serang, ID & Cibitung, ID & Cibitung, ID & & \\
        \cmidrule(lr){1-13}
        \multirow{8}{*}{IPinfo} & 1 & \texttt{194.242.34.201} & & 20612 & 0.00 & -- & -- & Zurich, CH & Zurich, CH & Glattbrugg, CH & \multirow{8}{*}{-0.81} & \multirow{8}{*}{0.97} \\
        & 2 & \texttt{91.206.52.37} & & 0 & 1.00 & -- & SwissIX, CH & -- & Zurich, CH & Zurich, CH & & \\
        & 3 & \texttt{82.220.13.18} & & 9044 & 1.00 & Zurich, CH & -- & Zurich, CH & Zurich, CH & Zurich, CH & & \\
        & 4 & \texttt{91.206.52.143} & & 0 & 0.00 & -- & SwissIX, CH & -- & Zurich, CH & Zurich, CH & & \\
        & 5 & \texttt{87.245.234.83} & & 9002 & 145.00 & -- & -- & Singapore, SG & Singapore, SG & Singapore, SG & & \\
        & 6 & \texttt{87.245.231.201} & & 9002 & 144.00 & -- & -- & Singapore, SG & Singapore, SG & Singapore, SG & & \\
        & 7 & \texttt{10.111.222.210} & \ding{51} & 0 & 145.00 & -- & -- & -- & -- & Singapore, SG & & \\
        & 10 & \texttt{206.237.97.100} & & 140443 & 158.00 & -- & -- & Serang, ID & Serang, ID & Cibitung, ID & & \\
        \cmidrule(lr){1-13}
        \multirow{8}{*}{MaxMind} & 1 & \texttt{194.242.34.201} & & 20612 & 0.00 & -- & -- & Zurich, CH & unknown, CH & Glattbrugg, CH & \multirow{8}{*}{-0.81} & \multirow{8}{*}{0.99} \\
        & 2 & \texttt{91.206.52.37} & & 0 & 1.00 & -- & SwissIX, CH & -- & unknown, CH & Zurich, CH & & \\
        & 3 & \texttt{82.220.13.18} & & 9044 & 1.00 & Zurich, CH & -- & Zurich, CH & Aarau, CH & Zurich, CH & & \\
        & 4 & \texttt{91.206.52.143} & & 0 & 0.00 & -- & SwissIX, CH & -- & unknown, CH & Zurich, CH & & \\
        & 5 & \texttt{87.245.234.83} & & 9002 & 145.00 & -- & -- & Singapore, SG & Singapore, SG & Singapore, SG & & \\
        & 6 & \texttt{87.245.231.201} & & 9002 & 144.00 & -- & -- & Singapore, SG & Singapore, SG & Singapore, SG & & \\
        & 7 & \texttt{10.111.222.210} & \ding{51} & 0 & 145.00 & -- & -- & -- & -- & Singapore, SG & & \\
        & 10 & \texttt{206.237.97.100} & & 140443 & 158.00 & -- & -- & Serang, ID & Cibitung, ID & Cibitung, ID & & \\
        \bottomrule
    \end{tabular}%
    }
\end{table*}

\begin{table*}[!t]
    \centering
    \caption{Traceroute case studies, continued. Hop RTT is the
    observed traceroute RTT in ms, Bogon is marked with \ding{51} only for
    bogon IPs, rDNS and IXP report public-source location evidence when
    available, and Validated loc. is the active validation location when
    available. Other public sources are omitted in these cases because they do
    not provide location mappings for the shown hops.}
    \label{tab:appendix:case-studies-continued}
    \tiny
    \setlength{\tabcolsep}{2.2pt}
    \renewcommand{\arraystretch}{0.86}
    \resizebox{\textwidth}{!}{%
    \begin{tabular}{@{}lclcrrlllllrr@{}}
        \toprule
        \textbf{Vendor} & \textbf{Hop} & \textbf{Hop IP} & \textbf{Bgn.} & \textbf{ASN} & \textbf{Hop RTT} & \textbf{rDNS} & \textbf{IXP} & \textbf{Validated loc.} & \textbf{GeoDB loc.} & \textbf{HMM loc.} & \textbf{PCS} & \textbf{$\mathcal{R}$} \\
        \midrule
        \multicolumn{13}{@{}l@{}}{\textbf{rDNS-supported correction.} Source anchor 1640; destination anchor 3936; measurement 79255204; timestamp 2025-08-01 00:04:25; hop 8 has Winnipeg rDNS and validation.} \\
        \midrule
        \multirow{5}{*}{DB-IP} & 2 & \texttt{93.92.99.40} & & 24586 & 2.00 & -- & -- & Amsterdam, NL & Amsterdam, NL & Rotterdam, NL & \multirow{5}{*}{-1.51} & \multirow{5}{*}{0.00} \\
        & 3 & \texttt{80.249.209.150} & & 204457 & 2.00 & -- & AMS-IX, NL & -- & Amsterdam (Amsterdam-Centrum), NL & Rotterdam, NL & & \\
        & 8 & \texttt{184.105.64.102} & & 6939 & 105.00 & Winnipeg, CA & -- & Winnipeg, CA & Singapore, SG & Winnipeg, CA & & \\
        & 9 & \texttt{184.105.34.255} & & 6939 & 108.00 & -- & -- & Karlstad, US & Montreal, CA & Karlstad, US & & \\
        & 10 & \texttt{69.89.207.87} & & 33362 & 108.00 & -- & -- & Karlstad, US & Karlstad, US & Karlstad, US & & \\
        \cmidrule(lr){1-13}
        \multirow{5}{*}{IP2Location} & 2 & \texttt{93.92.99.40} & & 24586 & 2.00 & -- & -- & Amsterdam, NL & Rotterdam, NL & Rotterdam, NL & \multirow{5}{*}{-1.51} & \multirow{5}{*}{0.00} \\
        & 3 & \texttt{80.249.209.150} & & 204457 & 2.00 & -- & AMS-IX, NL & -- & Amsterdam, NL & Rotterdam, NL & & \\
        & 8 & \texttt{184.105.64.102} & & 6939 & 105.00 & Winnipeg, CA & -- & Winnipeg, CA & Fremont, US & Winnipeg, CA & & \\
        & 9 & \texttt{184.105.34.255} & & 6939 & 108.00 & -- & -- & Karlstad, US & Wheeling, US & Karlstad, US & & \\
        & 10 & \texttt{69.89.207.87} & & 33362 & 108.00 & -- & -- & Karlstad, US & Karlstad, US & Karlstad, US & & \\
        \cmidrule(lr){1-13}
        \multirow{5}{*}{IPinfo} & 2 & \texttt{93.92.99.40} & & 24586 & 2.00 & -- & -- & Amsterdam, NL & Amsterdam, NL & Rotterdam, NL & \multirow{5}{*}{-1.51} & \multirow{5}{*}{0.14} \\
        & 3 & \texttt{80.249.209.150} & & 204457 & 2.00 & -- & AMS-IX, NL & -- & Amsterdam, NL & Rotterdam, NL & & \\
        & 8 & \texttt{184.105.64.102} & & 6939 & 105.00 & Winnipeg, CA & -- & Winnipeg, CA & Minneapolis, US & Winnipeg, CA & & \\
        & 9 & \texttt{184.105.34.255} & & 6939 & 108.00 & -- & -- & Karlstad, US & Fargo, US & Karlstad, US & & \\
        & 10 & \texttt{69.89.207.87} & & 33362 & 108.00 & -- & -- & Karlstad, US & Karlstad, US & Karlstad, US & & \\
        \cmidrule(lr){1-13}
        \multirow{5}{*}{MaxMind} & 2 & \texttt{93.92.99.40} & & 24586 & 2.00 & -- & -- & Amsterdam, NL & Den Hoorn, NL & Rotterdam, NL & \multirow{5}{*}{-1.51} & \multirow{5}{*}{0.33} \\
        & 3 & \texttt{80.249.209.150} & & 204457 & 2.00 & -- & AMS-IX, NL & -- & unknown, NL & Rotterdam, NL & & \\
        & 8 & \texttt{184.105.64.102} & & 6939 & 105.00 & Winnipeg, CA & -- & Winnipeg, CA & unknown, US & Winnipeg, CA & & \\
        & 9 & \texttt{184.105.34.255} & & 6939 & 108.00 & -- & -- & Karlstad, US & Minneapolis, US & Karlstad, US & & \\
        & 10 & \texttt{69.89.207.87} & & 33362 & 108.00 & -- & -- & Karlstad, US & Karlstad, US & Karlstad, US & & \\
        \midrule
        \multicolumn{13}{@{}l@{}}{\textbf{Potential MPLS tunnel.} Source anchor 3791; destination anchor 2084; measurement 23873216; timestamp 2025-08-01 00:04:48; hop 4 validates to Miami but HMM moves to Sydney.} \\
        \midrule
        \multirow{3}{*}{DB-IP}
            & 1 & \texttt{74.117.24.254} & & 32270 & 0.00 & -- & -- & Miami, US & Coral Gables (Douglas), US & Miami, US & \multirow{3}{*}{-1.63} & \multirow{3}{*}{0.00} \\
            & 4 & \texttt{66.110.9.65} & & 6453 & 231.00 & -- & -- & Miami, US & Miami, US & Sydney, AU & & \\
            & 255 & \texttt{139.99.219.16} & & 16276 & 272.00 & -- & -- & Sydney, AU & North Sydney, AU & Sydney, AU & & \\
        \cmidrule(lr){1-13}
        \multirow{3}{*}{IP2Location}
            & 1 & \texttt{74.117.24.254} & & 32270 & 0.00 & -- & -- & Miami, US & Miami, US & Miami, US & \multirow{3}{*}{-1.63} & \multirow{3}{*}{0.00} \\
            & 4 & \texttt{66.110.9.65} & & 6453 & 231.00 & -- & -- & Miami, US & Miami, US & Sydney, AU & & \\
            & 255 & \texttt{139.99.219.16} & & 16276 & 272.00 & -- & -- & Sydney, AU & Sydney, AU & Sydney, AU & & \\
        \cmidrule(lr){1-13}
        \multirow{3}{*}{IPinfo}
            & 1 & \texttt{74.117.24.254} & & 32270 & 0.00 & -- & -- & Miami, US & Miami, US & Miami, US & \multirow{3}{*}{-1.63} & \multirow{3}{*}{0.00} \\
            & 4 & \texttt{66.110.9.65} & & 6453 & 231.00 & -- & -- & Miami, US & Miami, US & Sydney, AU & & \\
            & 255 & \texttt{139.99.219.16} & & 16276 & 272.00 & -- & -- & Sydney, AU & Sydney, AU & Sydney, AU & & \\
        \cmidrule(lr){1-13}
        \multirow{3}{*}{MaxMind}
            & 1 & \texttt{74.117.24.254} & & 32270 & 0.00 & -- & -- & Miami, US & unknown, US & Miami, US & \multirow{3}{*}{-1.63} & \multirow{3}{*}{0.33} \\
            & 4 & \texttt{66.110.9.65} & & 6453 & 231.00 & -- & -- & Miami, US & Miami, US & Sydney, AU & & \\
            & 255 & \texttt{139.99.219.16} & & 16276 & 272.00 & -- & -- & Sydney, AU & Sydney, AU & Sydney, AU & & \\
        \bottomrule
    \end{tabular}%
    }
\end{table*}

\end{document}